\documentclass[preprint,aps,nofootinbib]{revtex4}
\input{epsf.tex}
\usepackage[dvips]{graphics,epsfig}
\def\be{\begin{equation}}
\def\ee{\end{equation}}
\def\ba{\begin{eqnarray}}
\def\ea{\end{eqnarray}}

\def\gtorder{\mathrel{\raise.3ex\hbox{$>$}\mkern-14mu
             \lower0.6ex\hbox{$\sim$}}}
\def\ltorder{\mathrel{\raise.3ex\hbox{$<$}\mkern-14mu
             \lower0.6ex\hbox{$\sim$}}}

\def\dalemb#1#2{{\vbox{\hrule height.#2pt
        \hbox{\vrule width.#2pt height#1pt \kern#1pt \vrule width.#2pt}
        \hrule height.#2pt}}}

\begin{document}

\rightline{DF/IST-5.2007}

\vskip 1cm
\begin{center}
{\bf{\large Model for Osteosarcoma-9 as a  Potent Factor 
in Cell Survival and Resistance to Apoptosis
}}
\vskip 0.5cm

Ekaterini Vourvouhaki${}^{1}$ \thanks{E-mail: kvourvouhaki@pasteur.gr}, 
Carla Carvalho${}^{2}$ \thanks{E-mail: ccarvalho@ist.edu},
Paulo Aguiar${}^{3}$ \thanks{E-mail: paguiar@ibmc.up.pt}\\

${}^1$ Hellenic Pasteur Institute, 127~Vasilissis Sofias Ave.,  
115~21 Athens\\
${}^2$ Departamento de F\'isica, Instituto Superior
T\'ecnico, Av.~Rovisco Pais, 
1049--001 Lisbon\\
${}^3$ Instituto de Biologia Molecular Celular, 
Rua Campo Alegre 823, 4150--180 Porto \\  

\vskip 1cm

{\large Abstract}
\end{center}

{\small
In this paper we use a simple model to explore the function 
of the gene Osteosarcoma-9. We are in particular interested in
understanding the role of this gene 
as a potent anti-apoptotic factor. 
The theoretical 
description is constrained by experimental data from
induction of apoptosis in cells where OS-9 is overexpressed. 
The data available suggest that OS-9 promotes cell viability and
confers resistance to apoptosis, potentially implicating OS-9 in the
survival of cancer cells.  
Three different apoptosis-inducing mechanisms were tested and are here
modeled.
A more complex and realistic model is also discussed. 
}


\vskip 1cm

\noindent${}^2$Corresponding author. \\
E-mail: ccarvalho@ist.edu\\
Phone: \hspace{0.cm}+351 21 8419077 \\
Fax:\hspace{0.7cm}+351 21 8419013\\

\newpage
\section{Introduction}


Of the various Expressed Sequence Tags 
of unknown function which appear to be regulated
when apoptosis is initiated \cite{1}, one was identified as the
gene Osteosarcoma-9 (OS-9 GI:580310), which is 
mapped to the chromosomal region 12q13-q15.
This region is amplified in various human cancers, including
sarcomas and osteosarcomas. 
The observation that none of the known genes in this region
is amplified in all tested tumors suggests that other genes
must exist that are commonly amplified in different cancer
types. The functional analysis of the OS-9 gene is thus of interest for its
potential anti-apoptotic role in the development of cancer.

However, little functional data exist for OS-9.
The identified sequence homology of OS-9 to both nucleolin and
protein tyrosine phosphatase 1B (PTP-1B) was found to entail 
no functional similarities \cite{8}. 
Being driven from a TATA-less promoter \cite{11}, however, suggests 
that OS-9 may be
involved in cell-cycle regulated housekeeping or cell viability \cite{8}.   
More recently, 
it was reported that OS-9 promotes oxygen-dependent degradation of
hypoxia inducible factor $1\alpha$ (HIF-$1\alpha$) via binding to both
HIF-1$\alpha$ and the HIF
prolyl hydroxylases  \cite{archenemy1}. This is a surprising result since
overexpression of HIF-1$\alpha$ is characteristic of most human cancer
and inhibition of HIF-1 impairs tumor growth.
A role for OS-9 in the control of tumor progression as a
tumor-suppressor gene has also been advanced \cite{archenemy2}.


In Ref.~\cite{os9} 
the first 
observational data set 
on a preliminary 
functional analysis of the product of the OS-9 gene was presented. 
The authors used overexpression of OS-9 isoform 2 cDNA in
two murine 
FDC-P1 cell line populations, namely FDB-P1-17V (17V) and
FDB-P1-17VIII (17VIII), 
and siRNA depletion of OS-9 specific mRNA in 
human umbilical vein endothelial cells (HUVEC)
to study the role of OS-9 in induction of apoptosis by three
mechanisms:
interleukin 3 
(IL-3) deprivation of 
IL-3-dependent murine myeloid cells \cite{il3},
staurosporine treatment \cite{staurosporine1, staurosporine2}
and tumor necrosis factor $\alpha$ (TNF$\alpha$) treatment \cite{tnfalpha}.
The data on the number of viable and apoptotic cells in the two 
OS-9-overexpressed cell line populations were 
obtained from trypan blue exclusion analysis and from 
fluoresence-activated cell sorter (FACS) 
analysis after treatment with Annexin V and propidium iodide (PI).
From the comparison with the cell population 
having the vector confering the Haemagglutinin (HA) gene only
(negative control),
the data suggest the following results:
1) promotion of cell viability upon induction of apoptosis by
withdrawal of IL-3; 
2) promotion of cell resistance to apoptosis upon
induction of apoptosis by addition of staurosporine and TNF$\alpha;$
3) concentration dependence of resistance to apoptosis by addition of
   staurosporine;  
4) resistance to apoptosis, with HUVEC'S precipitating to apoptosis
upon blockage of the OS-9 function;
5) no effect on cell proliferation of any of the mechanisms. 

This paper proposes a 
mechanistic model for the OS-9 functional data
presented in Ref.~\cite{os9}, thus giving theoretical support to the
interpretation of the experimental results by E.~Vourvouhaki et al. 
The model is general enough to avoid
speculations about the particular dynamics associated with OS-9,
but informative to the point of showing, in a quantitative manner, the
impact of OS-9 in cell fate. 
In particular, such a model must reproduce induction of apoptosis upon
treatment with 
an apoptotic agent or upon withdrawal of a vital agent, and explain
resistance to apoptosis and cell viability as a consequence of OS-9
overexpression.

This paper is organized as follows.
In section \ref{sec:model} we introduce the minimal model for the
kinetics of a population of murine FDC-P1 cells with OS-9 overexpressed
when subject to three different mechanisms for induction of apoptosis, namely
withdrawal of IL-3, addition of staurosporine and addition of TNF$\alpha.$
We analyse the biological system as a sequence of
biochemical interactions which we then formulate into a mechanistic
model encompassed by a system of kinetical equations.
In section \ref{sec:analysis} we optimize the parameters which
describe the experimental results by fitting to the data the kinetical
equations. 
For the optimal values for the parameters, the kinetical equations are
expected to model the biological system within the observational
errors and the limitations of a theoretical construct, as
discussed in section \ref{sec:discussion}.

\section{Model for OS-9 overexpression in apoptosis induction }
\label{sec:model}

In this section we analyse the biological system described by the
experimental data in Ref.~\cite{os9} as a sequence of biochemical
interactions modeled by the overexpression of the OS-9 protein in two
cell line populations of murine FDC-P1 cells, namely 17V and
17VIII. The level of OS-9 mRNA in 17VIII was confirmed by Real Time
 Polymerase Chain Reaction (PCR) 
to be approximately four times that of 17V \cite{os9}.   
Given the lack of knowledge on how the OS-9 protein acts in the cell,
the entailed biochemical interactions are reduced to the simplest
sensible functional form. In the biological system, we identify
two competing effects, namely induction of apoptosis on the one hand,
and on the other hand promotion of cell viability and resistance to
apoptosis due to OS-9 overexpression. 

Apoptosis is induced by the addition of an apoptotic agent, identified
as $a_{a},$ or by the withdrawal of a vital agent to the survival of
the cells, identified as $v_{a}$. Here we describe induction of
apoptosis 
by a decay rate of viable cells in the presence of an apoptotic agent
or absence of a vital agent. 
These rate parameters are characteristic of the apoptotic mechanism and
hence assumed to be dependent on the specific agent. 

The cell response to apoptosis induction will depend on the induction
mechanism. 
Thus, addition of an apoptotic agent will provoke the cell to resist
apoptosis, in which case we will be assessing the OS-9 action in 
resistance to apoptosis. Conversely, withdrawal of a vital agent will
challenge the cell to strive for survival, in which case we will be
assessing the action of OS-9 in promoting cell viability. 

These are the basic biochemical interactions which, together with the
normal viable cell growth, must be contemplated by any model that
attempts to describe the biological system being considered. We now
proceed to formulate the simplest possible kinetical model containing
just as many parameters as the data readily available allow us to
constrain. An extension of such model for an arbitrary number of
possible intermediary mechanical interceptions is presented as an
illustration of the increased level of data required to constrain an
increased level of interaction.

\subsection{Kinetical Model}

In our first attempt to model the OS9 role, we collapse the
biochemical interactions into a minimal set of kinetical equations. This
first-order approximation avoids speculation and allows us to
concentrate on the effects of OS-9 overexpression on promotion of
viability and resistance to apoptosis. In the simplest plausible
model, we have three variables: viable cells $c_{v},$ apoptotic cells
$c_{a}$ and dead cells $c_{n}.$
The evolution in time of the number of cells in each of these cell
stages is dependent on the reaction of the cell population to the
absence of vital agent $v_{a}$ or on the interaction with the
apoptotic agent $a_{a}$, and on the subsequent cell response derived from
an OS-9 overexpression.

The biochemical interactions can be analysed in fundamental
phenomenological processes which we represent schematically by the
following system of kinetical equations: 
\ba
c_{v} & \stackrel{\lambda}{\longrightarrow} & 2c_{v}, \label{growth}\\
c_{v} 
& \stackrel{\alpha(1-\beta)}{\longrightarrow} & c_{a}, \label{induction} \\
c_{a} & \stackrel{\gamma(1-\delta)}{\longrightarrow} & c_{n} \label{death} 
\label{}.
\ea
Here Eqn.~(\ref{growth}) corresponds to cell proliferation, 
Eqn.~(\ref{induction}) to apoptosis induction by either addition of an
apoptotic agent or withdrawal of a vital agent 
and  Eqn.~(\ref{death}) to cell death. 
The reactions will now be described in detail.

Viable cells grow at a rate $\lambda.$ The absence of any effect of
the OS-9 overexpression on cell proliferation, as concluded from the
DNA content analysis, implies that $\lambda =constant.$  The effective
cell growth, however, depends on the ratio of the number of viable
cells to the carrying capacity $c_{{v(max)}}$. As a first
approximation, we assume that the cells proliferate effectively at a
rate $\lambda(1 -c_{v}/c_{v(max)})$. The choice of the linear
functional form for the effective cell proliferation rate on the
carrying capacity was motivated by the treatment of various aspects of
cancer modeling discussed in References \cite{cbc, mm}.  

Equation (\ref{induction}) describes the transition from viable cells
to apoptotic cells induced by either addition of an apoptotic agent
$a_a$ or removal of a vital agent $v_a$. For both apoptosis induction
processes we use the same expression for the conversion rate: $\alpha
(1 -\beta).$ 
The two
parameters involved are used to segregate the effect of OS-9
overexpression. The parameter $\beta$ is a function of the OS-9
overexpresssion, $\beta =\beta$([OS-9]), and takes values in
the interval $[0,1]$. The parameter $\alpha$ absorbs all other
contributions to the conversion rate and depends on the specific agent
leading to apoptosis. All experimental situations described in this
paper use either one of these two induction processes; no experiment
uses both. Although both apoptosis-inducing processes have the same
rate expression, the interpretation given to the parameter $\beta$ is
different. 
In the case of vital agent withdrawal, the $\beta$ parameter of
Eqn.~(\ref{induction}) measures promotion of cell viability modulated
by the OS-9 overexpression. However, in the case of apoptotic agent
addition,
$\beta$ reflects resistance of a viable cell to becoming apoptotic. 
This process must thus take place upstream in the apoptotic signal
pathway before the activation of the initiator caspases
(caspase-8, -9 and -12) 
\cite{apoptosis}.  

Cell death, as expressed in Eqn.~(\ref{death}), progresses at a rate
given by the expression $\gamma (1 -\delta).$ 
As done previously, this expression segregates the dependence on OS-9
overexpression. The parameter $\delta$ is a function of the OS-9
overexpression, $\delta =\delta$([OS-9]), and takes values in
the interval $[0,1]$. Apoptotic cells will enter the final stage of
apoptosis at a rate $\gamma$. Correspondingly, the parameter $\delta$
reflects resistance of an apoptotic cell to entering such stage
characterized by the cell turning into an apoptotic body and
eventually dying. This effect will take place downstream in the
apoptotic signal pathway before the activation of the executioner
caspases (caspase-3, -6 and -7) 
\cite{apoptosis}.  
 
It is important to emphasise the modulation produced by OS-9
overexpression in these reactions. This modulation is encapsulated in
the parameters $\beta =\beta$([OS-9]) and $\delta
=\delta$([OS-9]), which are such that $\beta, \delta \in
[0,1],$ where $0$ denotes the control (normal) expression and all
other values reflect a contribution of OS-9 overexpression. 

We now proceed to formulate the phenomenological model into a system
of differential equations for the three variables. In the case of
induced apoptosis by the addition of one apoptotic agent, the
interaction can be broken down as follows: 
a) binding of $a_{a}$ to $c_{v};$ 
b) interception by the OS-9-codified protein of a signal pathway
triggered from an effective binding;  
c) blocking of apoptosis induction in $c_{v}$  or blocking of necrosis
induction in $c_{a},$ depending on the position in the signal pathway
of the interception;   
d) survival or death of $c_{a}.$ 

For a contact rate $\alpha,$ $\alpha$ molecules of the apoptotic agent
will bind to a cell per unit time.  
If we assume that from each contact point a signal for apoptosis is
always started, then $\alpha c_{v}$ will be the number of
triggered apoptotic signals per unit time.  
Before starting the upstream apoptosis-inducing caspase chain
reaction, $\beta (\alpha c_{v})$ apoptotic signals interact with
a molecule of the OS-9-codified protein per unit time, resulting in
the interruption of the signal pathway. The remaining $(1
-\beta)(\alpha c_{v})$ signals will successfully induce apoptosis.  
In cells where apoptosis has already been triggered, and in the
absence of OS-9 overexpression, the final stage of apoptosis is
reached at a rate $\gamma c_a.$ In the presence of OS-9
overexpression, however, for $\delta (\gamma c_a)$ successful OS-9-led
interruption of a downstream signal for necrosis, the final stage of
apoptosis will instead be reached at $\gamma c_{a}( 1 -\delta).$ 
Combining the reactions described above, we find that the variation in
time of the number of viable cells is given by  
\ba
{d c_{v}\over dt} 
=\lambda c_{v}\left( 1 -{c_{v}\over c_{v(max)}}\right) 
-{ {\alpha c_{v}}} ( 1 -\beta).
\label{eqn:cv}
\ea
Given the time scale of the experiment, cell decay is rendered
negligible and thus unaccounted for. This is supported by the data
recorded in the FACS diagrams  
in Ref.~\cite{os9}, which suggest that
necrosis is sourced by apoptosis rather than by decay.
It follows that the number of apoptotic cells will vary in time as
\ba
{d c_{a}\over dt}
={ {\alpha c_{v}}} ( 1 - \beta)
-\gamma c_{a}(1 -\delta),
\label{eqn:ca}
\ea
and the number of dead cells as
\ba
{d c_{n}\over dt}
=\gamma c_{a}(1 -\delta).
\label{eqn:cn}
\ea
This is an example of a quasispecies-type system \cite{cbc}.

In the case of induced apoptosis by  withdrawal of a
vital agent, the interaction can be broken down in a similar way to
the previous case, 
with the apoptosis signal now being driven through the 
mitochondria/cytochrome c pathway \cite{il3}. 
The induction of apoptosis can still be described by a decay mechanism 
of rate $\alpha c_{v}.$ This means that 
$\alpha$ apoptoptic signals are triggered per unit time in a cell
subject to a vital agent-depleted medium.
Of the total $\alpha c_{v}$ apoptotic signals generated, 
$\beta (\alpha c_{v})$ are intercepted by the OS-9
protein, which measures the rate of failure of apoptosis induction.

Both mechanisms discussed here are independent and each comprises
either addition of an apoptotic agent or withdrawal of a vital
agent. Consequently, we would expect that each highlights an
independent function of OS-9, respectively promotion of viability or
resistance to apoptosis.

\subsection{Chemical Model}

The model described in the previous subsection can be regarded as a
{\it black box} model which acts on a given initial state to reproduce
the observed final state in a OS-9-dependent way.
If the detailed {\it modus operandus} of the OS-9 action in
the cell was known, then the system would be approximated by a
{\it multiple little black box} model, with smaller black boxes
describing intermediate reactions triggered by intermediary apoptotic
agents and constrained by intermediary initial and final known states.
For one intermediary apoptotic agent, $a_{a}^{(1)},$ 
in the upstream pathway between the trigger
of the exterior apoptotic agent on the membrane of the cell and the
interception by the OS-9 protein which binds to the signal at the
rate $\alpha_{up (1)},$ the variation in time of the number
of viable cells will be given by
\ba
{d c_{v}\over dt}
&=&\lambda c_{v}\left( 1 -{c_{v}\over {c_{v(max)}}}\right)
-\alpha c_{v}\left[ 
\alpha_{up (1)}
-\beta \alpha_{up (1)}
-\left( 1 -\alpha_{up (1)}\right)\right]\cr
&=&\lambda c_{v}\left( 1 -{c_{v}\over {c_{v(max)}}}\right)
-\alpha c_{v}\left[ 
\alpha_{up (1)}\left( 2 -\beta\right) 
-1\right];
\ea
for two intermediary upstream apoptotic agents, $a_{a}^{(1)}$ and
$a_{a}^{(2)},$ 
with binding rates $\alpha_{up (1)}$ and $\alpha_{up (2)}$
respectively, we have that
\ba
{d c_{v}\over dt}
&=&\lambda c_{v}\left( 1 -{c_{v}\over {c_{v(max)}}}\right)
-\alpha c_{v}
 \alpha_{up (1)}
  \alpha_{up (2)}\cr
&+&\alpha c_{v}\left[ 
\beta 
 \alpha_{up (1)}
  \alpha_{up (2)} 
+\alpha_{up (1)}
 \left( 1-\alpha_{up (2)}\right)
+\left( 1 -\alpha_{up (1)}\right)\right]\cr
&=&\lambda c_{v}\left( 1 -{c_{v}\over {c_{v(max)}}}\right)
-\alpha c_{v}
\left[
\alpha_{up (1)}
 \alpha_{up (2)}
  \left( 2 -\beta\right) 
-1\right];
\ea
and similarly for $n$ intermediary chemical agents upstream in the
apoptic pathway
\ba
{d c_{v}\over dt}
&=&\lambda c_{v}\left( 1 -{c_{v}\over {c_{v(max)}}}\right)
-\alpha a_{a}c_{v}\left[
\left(2 -\beta\right)
 \prod^{n}_{i=1}\alpha_{up (i)}
-1\right].
\label{eqn:cv:n}
\ea
To account for the OS-9-modulated interception of the downstream
signal pathway we must allow for intermediary downstream necrotic agents,
$n_{a} ^{(n)},$ binding to the OS-9 protein at rates $\alpha_{down
  (n)}.$ It follows that 
\ba
{d c_{a}\over dt}
=\alpha a_{a}c_{v}\left[
\left(2 -\beta\right)
 \prod^{n}_{i=1}\alpha_{up (i)} 
-1\right]
-\gamma c_{a} \left[
\left(2 -\delta\right)
 \prod^{n}_{i=1}\alpha_{down (i)}
-1\right].
\label{eqn:ca:n}
\ea
The role of each
of these agents in the transmission of the apoptotic signal must be
known and included in the dynamical system. To study the potential
role of such a weakly known gene such as the OS-9, this model is
therefore inadequate. 
As the logical extension of the kinetical model to a series of such
smaller systems, this model could be applied to a multiple level
interaction approach to biological systems with available multiple level data
constraints \cite{ref}. It might also provide further insight into some
unexpected results obtained with the kinetical model. We hope to
pursue this in the future.

\section{Experimental and Theoretical Results}
\label{sec:analysis}

In this section we proceed to use the system of first-order, coupled
differential equations derived in the previous section to model the
data reported in Ref.~\cite{os9}. This system is used to model the two
different apoptotic inducing mechanisms, namely addition of an
apoptotic agent or withdrawal of a vital agent. 
In this paper we shall only use what we called the kinetic model.
We want to solve for the time evolution of the number of viable cells
$c_{v}(t)$ and of apoptotic cells $c_{a}(t)$ using the data in
Ref.~\cite{os9} to constrain the parameters $\alpha,$ $\beta,$
$\gamma$ and $\delta$ of the differential equations. In particular we
are interested in the values for the parameters $\beta$ and $\delta$
which measure the contribution of the OS-9 overexpression in the
system. 

The data consist of the average of nine readings of 
the number of viable cells and the number of apoptotic cells 
obtained from trypan blue exclusion analysis and from 
FACS analysis after treatment with Annexin V and PI,
all experiments having been performed three times and each in triplicate.
The FACS analysis 
sorts cells that are
1) alive,
2) undergoing apoptosis, 
3) undergoing necrosis and 
that are 4) in between apoptosis and necrosis, thus allowing to
discriminate alive, apoptotic and dead cells.
The trypan blue exclusion analysis simply separates dead from alive
cells, lacking the 
sophistication and sensitivity to discriminate viable cells from
apoptotic cells. 
In a modeling study that considers all three of the cell stages, the
data from the typan blue exclusion analysis are inappropriate for use and
hence will not be included here.

The solutions were determined numerically with {\sc mathematica}
(Wolfram Research).  
For the parameter estimation we implemented a sampling algorithm which 
calculates a cost function of a list of values for each
parameter that we want to estimate. The cost function is defined as
the sum of the square of differences between the experimental points and
the corresponding points of the solution to the kinetical system. The
minimum of the cost function line, or surface, defines the parameter
value which optimally fits the experimental data. The precision of the
method is defined by the density of points in the search space. 
The optimal values thus determined for all the
parameters were then used to plot the solutions for $c_{v}(t)$ and
$c_{a}(t)$ which we present in this paper.
We applied the algorithm to both the average values and the
corresponding maxima and minima encompassed by the error bars. For
each data set, we obtained from three to eight estimates 
for (and depending on) each parameter by 
fitting the equations to the average and to the extreme data points. 
The parameters obtained from the average data points were combined to
produce the {\it induced average} solution 
(represented by a continuous line); 
the parameters obtained from the extreme data points were
combined to produce the maximum and minimum solutions 
(represented by dashed lines siding the continuous line), which
reflect the propagation of the measurement errors and thus define the 
{\it induced error interval} of all possible curves capable of optimally
fitting the data within the experimental errors.
All the curves show percentages normalized to the initial
number of viable cells, $c_{v}(t=0) =1\times 10^{5}.$

From the experimental curves we can extrapolate the values for
$\lambda$ and $c_{v(max)}$ adjusted to the system. Here we use
$\lambda =0.01$ and $c_{v(max)} =20\times 10^{5}.$ 
From the data on the HA-only marked (negative control) population we
estimate the parameters which do not depend on the OS-9
overexpression, namely $\alpha$ and $\gamma.$ 
We proceed to apply them to the OS-9-overexpressed cell line
populations to estimate the parameters which measure the overexpression of
OS-9, namely $\beta$ and $\delta.$
The data analysis of each apoptosis-inducing mechanism is introduced
by the interpretation advanced in Ref.~\cite{os9} based on a
qualitative reading of the experimental results, which we then
confront with the
interpretation of the quantified, data-constrained parameters of the model.

\subsection{Withdrawal of IL-3 in FDC-P1 cells}

\begin{figure}[h]
\setlength{\unitlength}{1cm}
\begin{center}
\begin{minipage}[t]{8.cm}
\begin{picture}(8.,8.)
\centerline 
{\hbox{\psfig{file=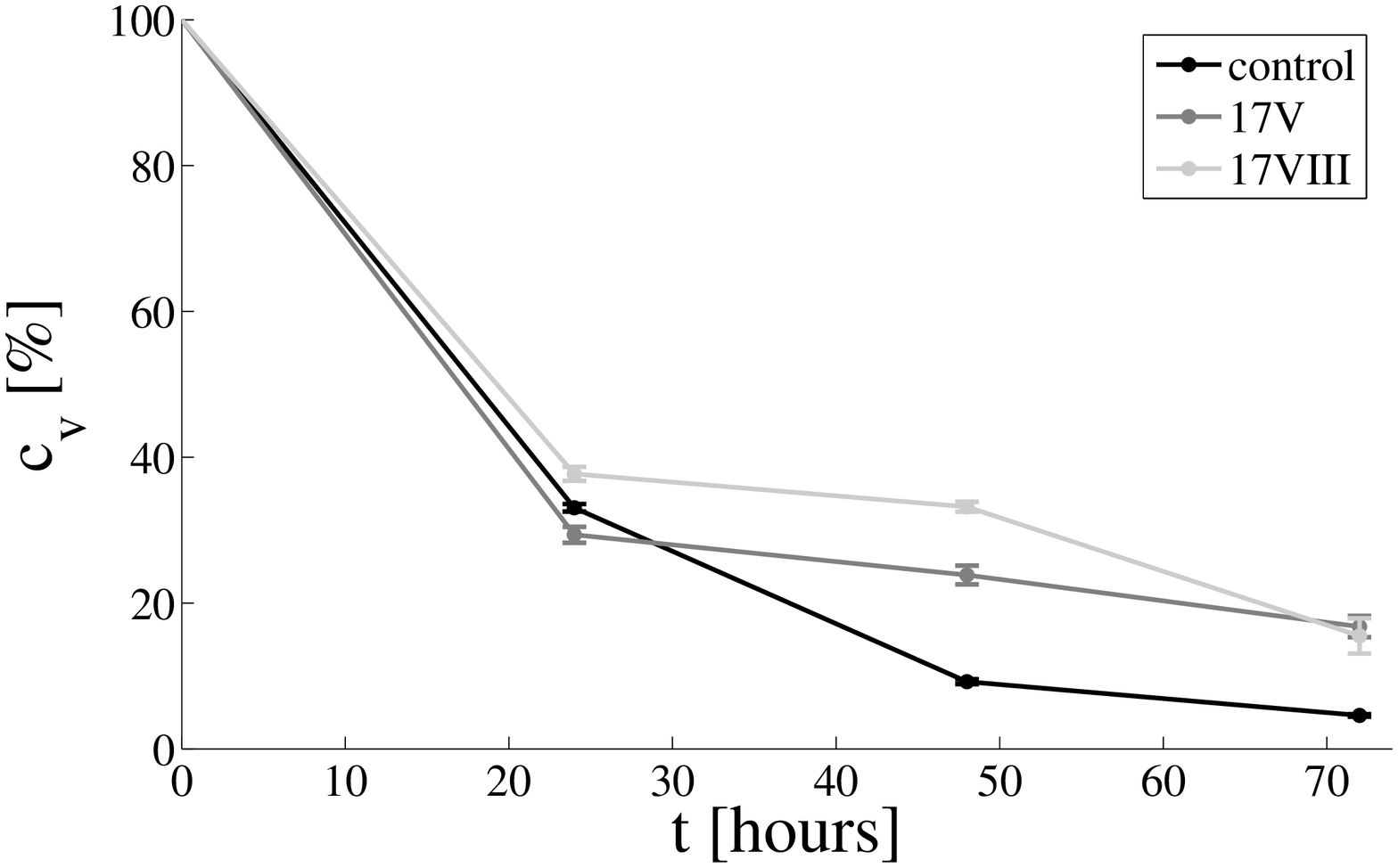,height=5.7cm}}}
\put(-4,6.0){$(a)$}
\end{picture}\par
\end{minipage}
\hfill
\begin{minipage}[t]{8.cm}
\begin{picture}(8.,8.)
\centerline 
{\hbox{\psfig{file=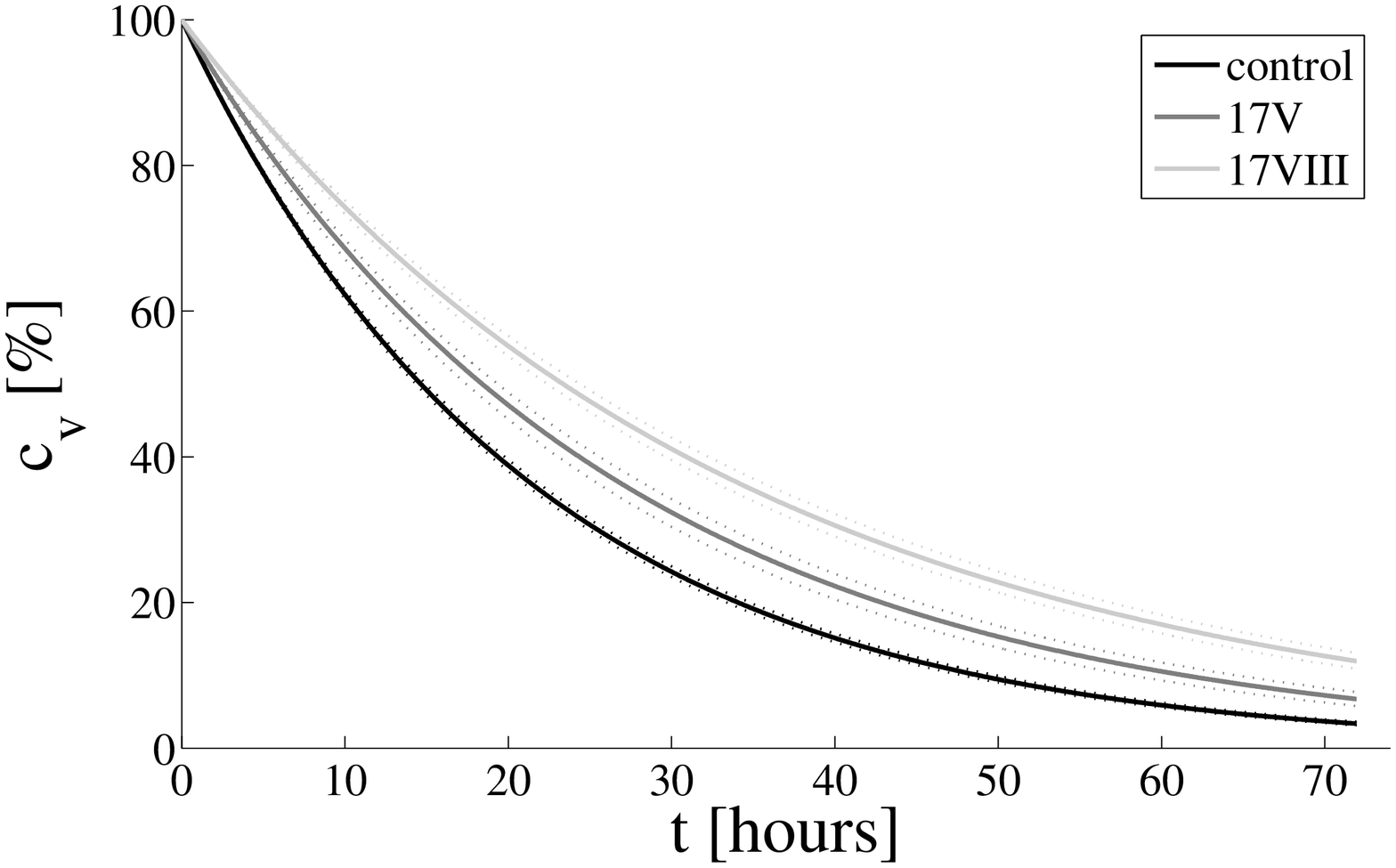,height=5.7cm}}}
\put(-4,6.0){$(b)$}
\end{picture}\par
\end{minipage}
\end{center}
\caption{\small \baselineskip=4pt {
{\bf Number of viable cells after withdrawal of IL-3,
in the presence and in the absence of OS-9, 
from FACS analysis.}
}}
\label{fig:data:cv:2b}
\end{figure}

\begin{figure}[h]
\setlength{\unitlength}{1cm}
\begin{center}
\begin{minipage}[t]{8.cm}
\begin{picture}(8.,8.)
\centerline 
{\hbox{\psfig{file=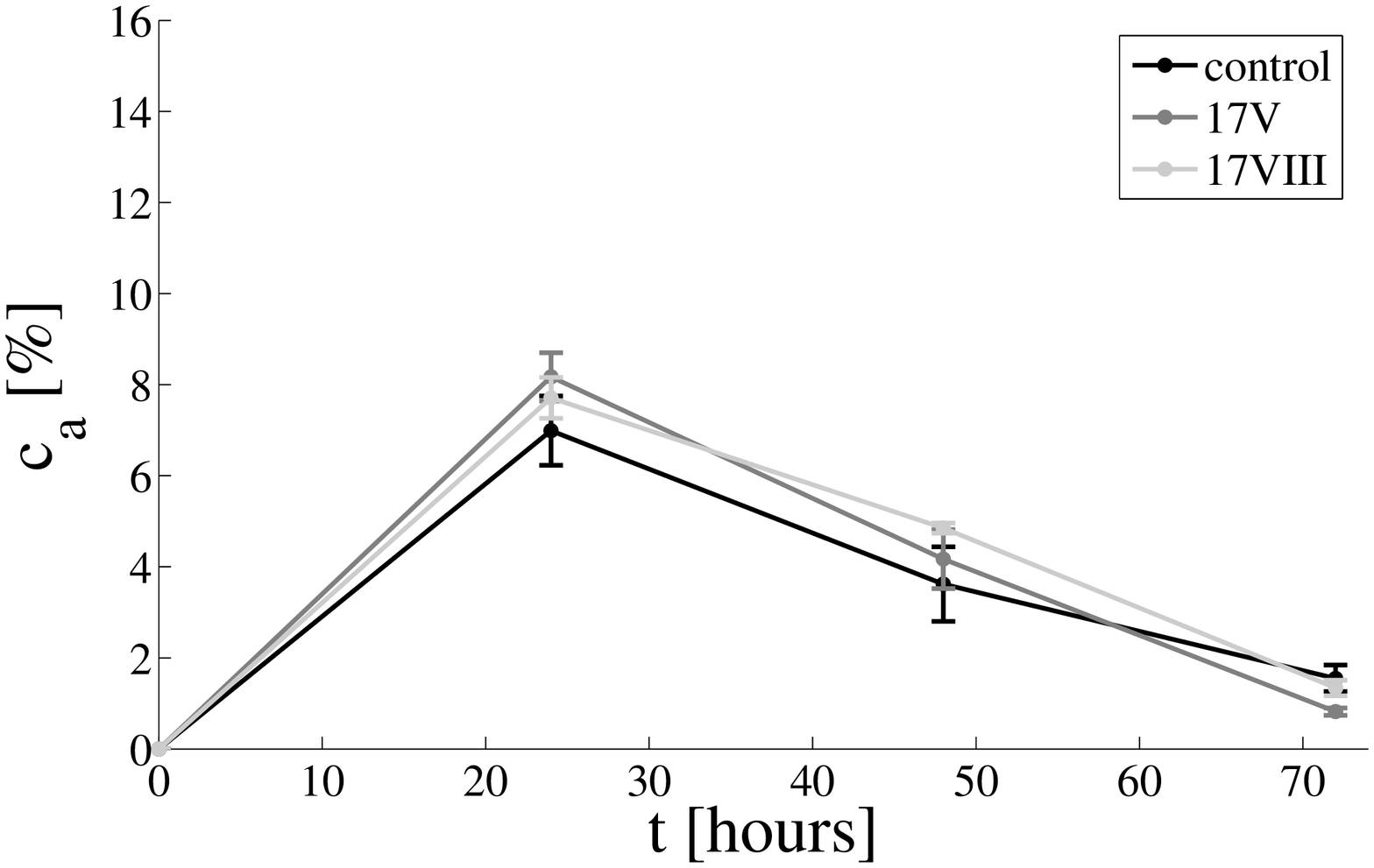,height=5.7cm}}}
\put(-4,6.0){$(a)$}
\end{picture}\par
\end{minipage}
\hfill
\begin{minipage}[t]{8.cm}
\begin{picture}(8.,8.)
\centerline 
{\hbox{\psfig{file=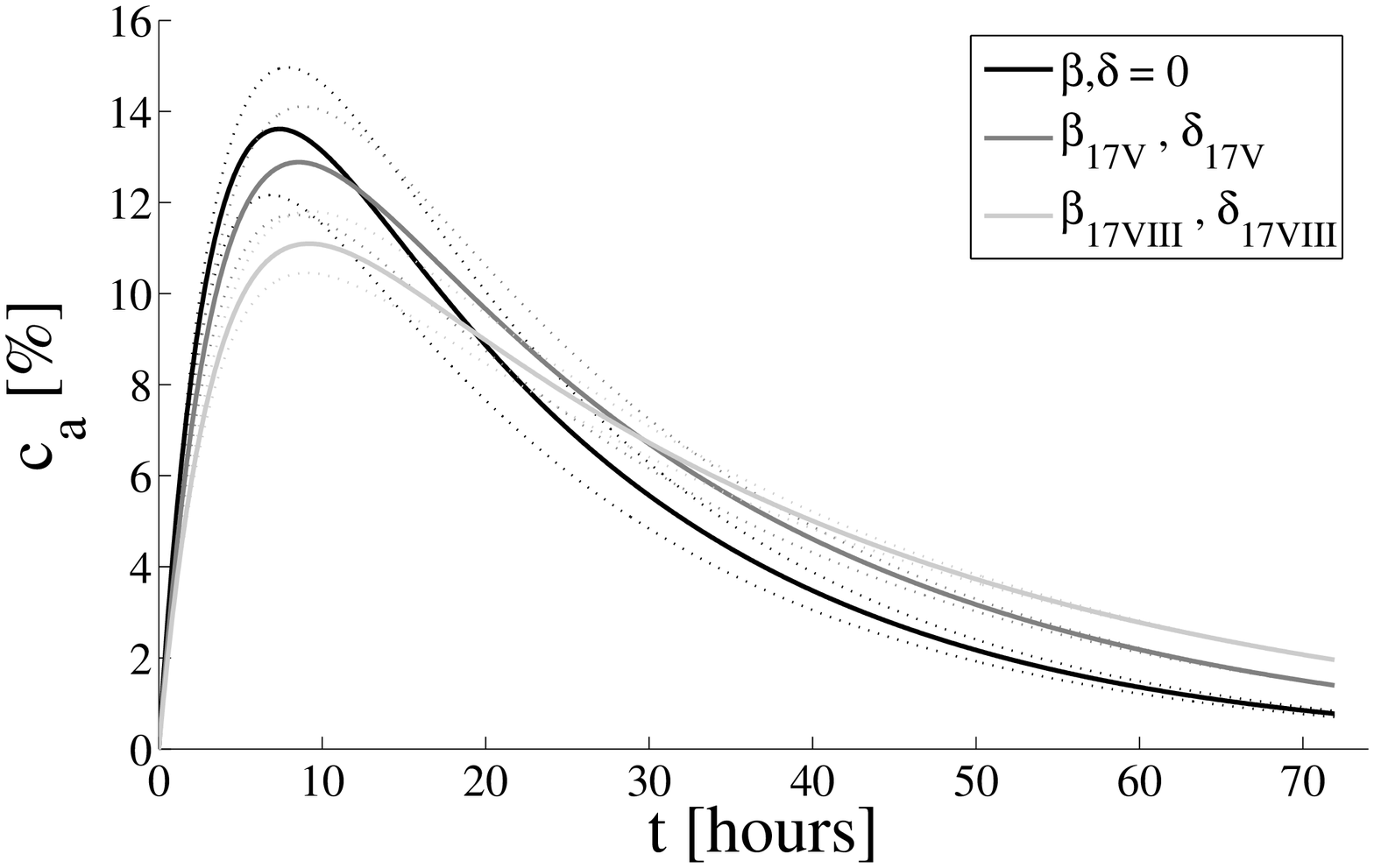,height=5.7cm}}}
\put(-4,6.0){$(b)$}
\end{picture}\par
\end{minipage}
\end{center}
\caption{\small \baselineskip=4pt {
{\bf Number of apoptotic cells after withdrawal of IL-3, 
in the presence and in the absence of OS-9, 
from FACS analysis.}
}}
\label{fig:data:ca:2c}
\end{figure}

Here we analyse apoptosis induction by IL-3 deprivation
of IL-3-dependent cells.
The corresponding apoptotic mechanism is that of withdrawal of a vital
agent and is described by the system of equations
(\ref{eqn:cv}--\ref{eqn:cn}), with $\beta$ having the
interpretation of OS-9 overexpression modulation in cell viability. 
The experimental results are collected 
in Figs.~2B and 2C of Ref.~\cite{os9}, which we 
reproduce in Figs.~\ref{fig:data:cv:2b}(a) and \ref{fig:data:ca:2c}(a)
respectively.  

First we solve for $c_{v}(t)$ given by  Eqn.~(\ref{eqn:cv}) and
fit it to the number of viable cells shown  
in Fig.~\ref{fig:data:cv:2b}(a).
In the absence of OS-9 overexpression this equation has only one free
parameter, namely $\alpha.$ Setting $\beta =0,$ we fit the solution
to the curve for the number of viable cells 
in the HA-only marked cell population
to find $\alpha =0.057^{+0.001}_{-0.001}.$
In the presence of OS-9 overexpression we have additionally the
parameter $\beta$ to 
consider. We now use the results found for $\alpha$ to fit $c_{v}(t)$
to the experimental curve for the number of viable cells 
in the OS-9-overexpressed cell line 
populations and thus to find the corresponding $\beta.$  
We obtain 
$\beta_{17V} =0.17^{+0.04}_{-0.05}$ 
and $\beta_{17VIII} =0.31^{+0.03}_{-0.04}$  
for the 17V and the 17VIII cell line populations respectively.  
The resulting plots are depicted in 
Fig.~\ref{fig:data:cv:2b}(b). 

We then solve for $c_{a}(t)$ given by Eqn.~(\ref{eqn:ca})
and fit it to the number of apoptotic cells shown in 
Fig.~\ref{fig:data:ca:2c}(a) 
to determine the parameters $\gamma$ and $\delta.$ 
Using $\alpha =0.057^{+0.001}_{-0.001}$ and setting $\beta, \delta =0,$
the fitting of the solution to the number of apoptotic cells 
in the negative control population 
gives $\gamma =0.30^{+0.04}_{-0.04}.$ 
Substituting now also for $\beta$ the results 
$\beta_{17V} =0.17^{+0.04}_{-0.05}$ and
$\beta_{17VIII} =0.31^{+0.03}_{-0.04},$ 
the fitting of the solution to the 
number of apoptotic cells 
in the OS-9-overexpressed cell line
populations gives 
$\delta_{17V}=0.099^{+0.16}_{-0.099}$ 
and $\delta_{17VIII}=0.086^{+0.14}_{-0.086}$ respectively. 
The resulting plots are depicted in Fig.~\ref{fig:data:ca:2c}(b).


In accordance with the experimental data, 
our mathematical model returns no significant difference between the
number of apoptotic cells in the negative control population and that
in the OS-9-overexpressed cell line populations, as illustrated by the
shared pattern of the curves. We note that it captures rather closely the
crossing point where the number of apoptotic cells in the 
OS-9-overexpressed cell populations is the same and the starting dominance
of the 17V population over the 17VIII population is reversed.
As anticipated by E.~Vourvouhaki et al. \cite{os9}, this case
illustrates promotion of cell viability (as shown by $\beta
\not=0$) and possibly downstream resistance to apoptosis (as shown by $\delta
\not=0$). 
Our mathematical model allows to separate the effect of
promotion of viability, 
as encapsulated in a non-vanishing $\beta,$ from the effect of
downstream resistance to apoptosis, as encapsulated in a non-vanishing
$\delta.$

\subsection{Staurosporine treatment in FDC-P1 cells}

Here we analyse apoptosis induction by treatment with staurosporine.
The apoptotic mechanism is that of addition of an apoptotic agent and
is described by the system of equations
(\ref{eqn:cv}--\ref{eqn:cn}),  with $\beta$ having the interpretation
of OS-9 overexpression modulation in apoptosis resistance. 
The experimental results
are collected in Figs.~5B and 5C of Ref.~\cite{os9}.

Two different concentrations of staurosporine were considered: 
$1\mu M$ and $0.1\mu M.$
The data for [staurosporine]=$1\mu M$ are reproduced 
in Figs.~\ref{fig:data:cv:5b:1muM}(a) and \ref{fig:data:ca:5c:1muM}(a);
the data for [staurosporine]=$0.1\mu M$ are reproduced
in Figs.~\ref{fig:data:cv:5b:0.1muM}(a) and \ref{fig:data:ca:5c:0.1muM}(a).
We do not impose any additional constrain in the system to account for
the concentration change. In other words, we let the parameters
$\alpha$ and $\beta$ absorb the changes induced by the different
concentrations. 


\subsubsection{[Staurosporine]=$1\mu M$}

\begin{figure}[t]
\setlength{\unitlength}{1cm}
\begin{center}
\begin{minipage}[t]{8.cm}
\begin{picture}(8.,8.)
\centerline 
{\hbox{\psfig{file=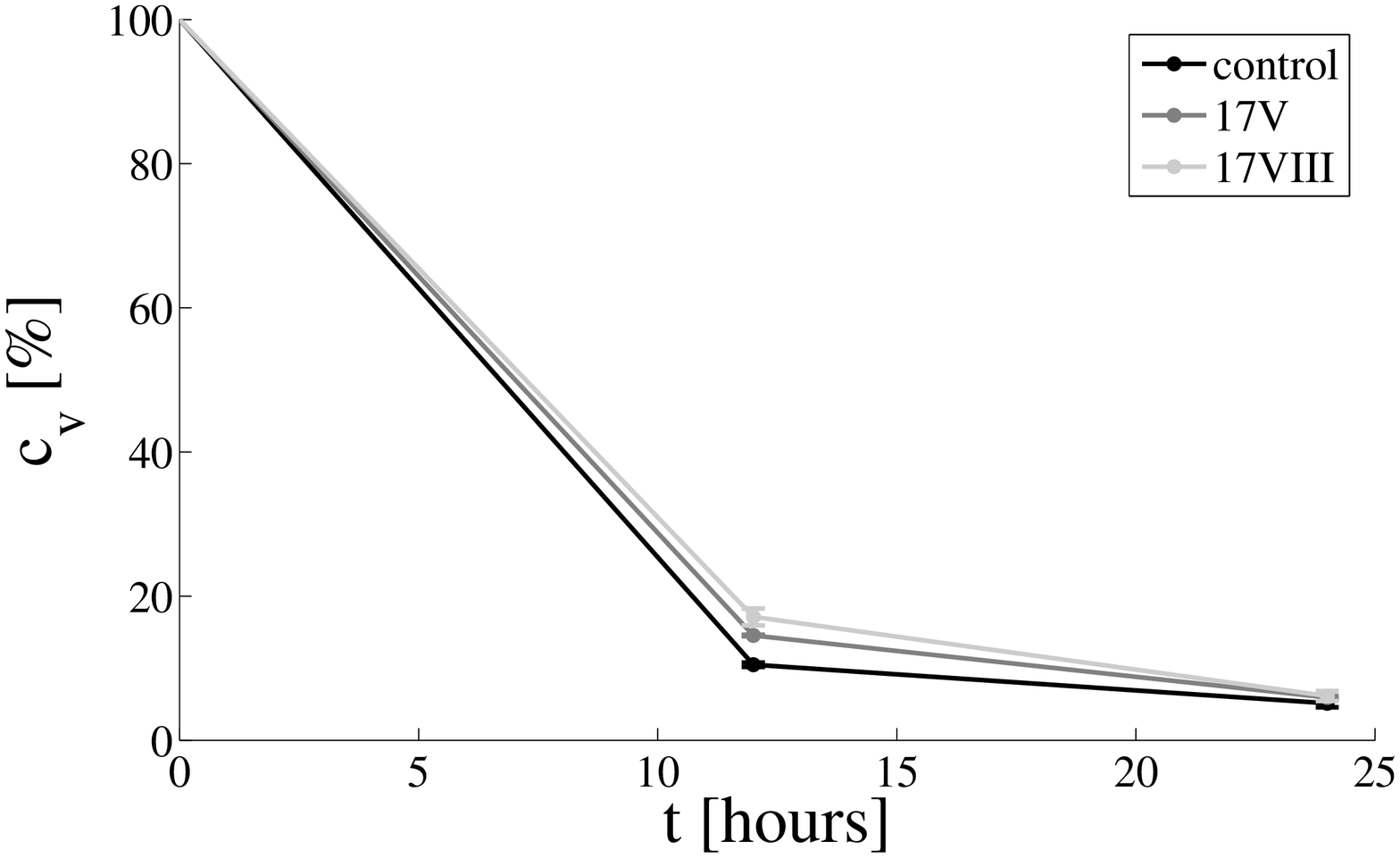,height=5.7cm}}}
\put(-4,6.0){$(a)$}
\end{picture}\par
\end{minipage}
\hfill
\begin{minipage}[t]{8.cm}
\begin{picture}(8.,8.)
\centerline 
{\hbox{\psfig{file=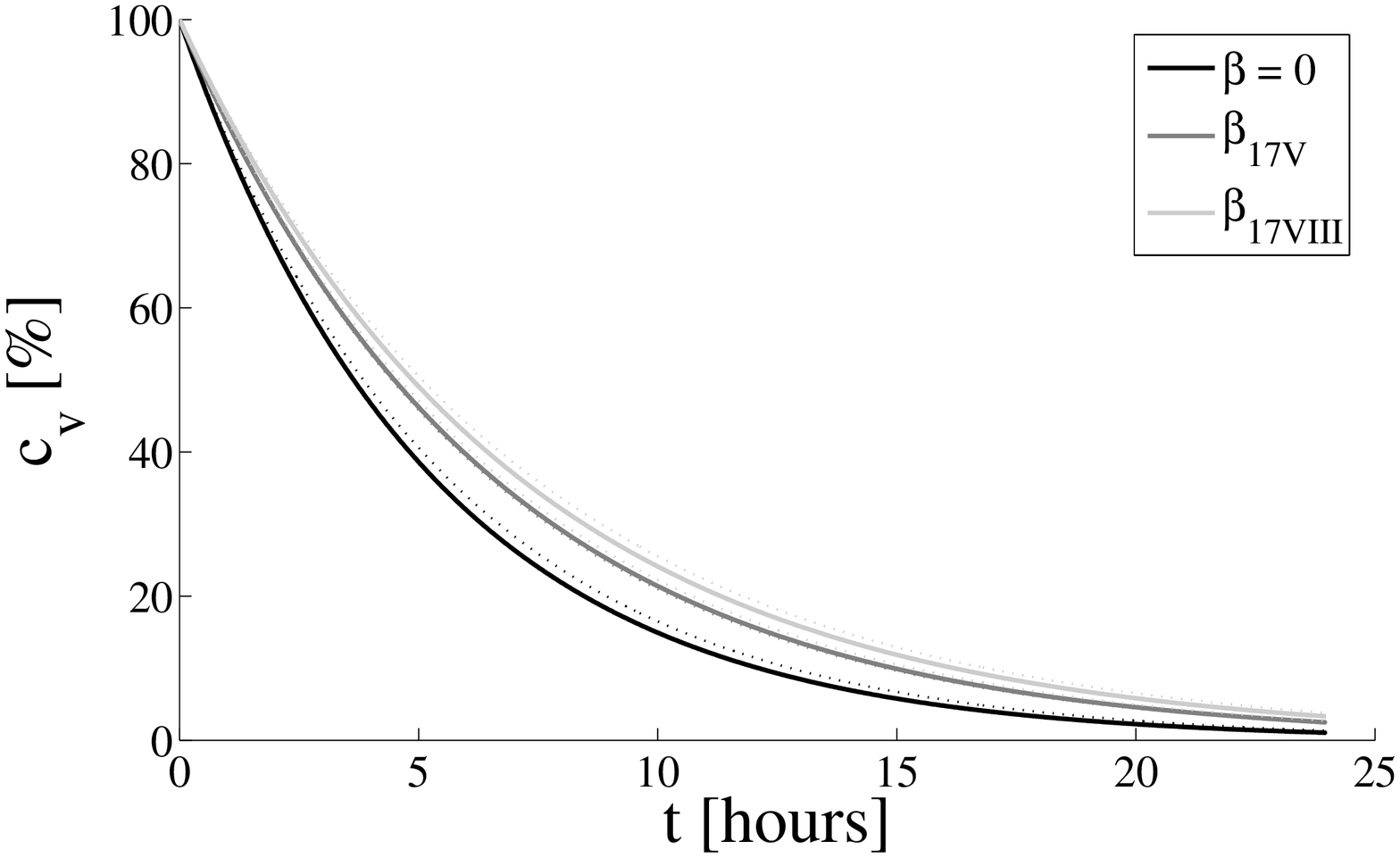,height=5.7cm}}}
\put(-4,6.0){$(b)$}
\end{picture}\par
\end{minipage}
\end{center}
\caption{\small \baselineskip=4pt {
{\bf Number of viable cells after treatment with
  staurosporine $1\mu M,$ in the presence and in the absence of OS9,
  from FACS analysis.}
}}
\label{fig:data:cv:5b:1muM}
\end{figure}

First we solve for $c_{v}(t)$ for [staurosporine]=$1\mu M,$  given by
Eqn.~(\ref{eqn:cv}), and fit it to 
Fig.~\ref{fig:data:cv:5b:1muM}(a)
to find $\alpha =0.20^{+0.005}_{-0.01}$  
from $c_{v}$ in the negative control population, 
and $\beta _{17V} =0.18^{+0.005}_{-0.05}$ 
and $\beta _{17VIII} =0.24^{+0.03}_{-0.07}$  
from the corresponding OS-9-overexpressed cell line 
populations. The results are plotted in
Fig.~\ref{fig:data:cv:5b:1muM}(b).

\begin{figure}[p]
\setlength{\unitlength}{1cm}
\begin{center}
\begin{minipage}[t]{8.cm}
\begin{picture}(8.,8.)
\centerline 
{\hbox{\psfig{file=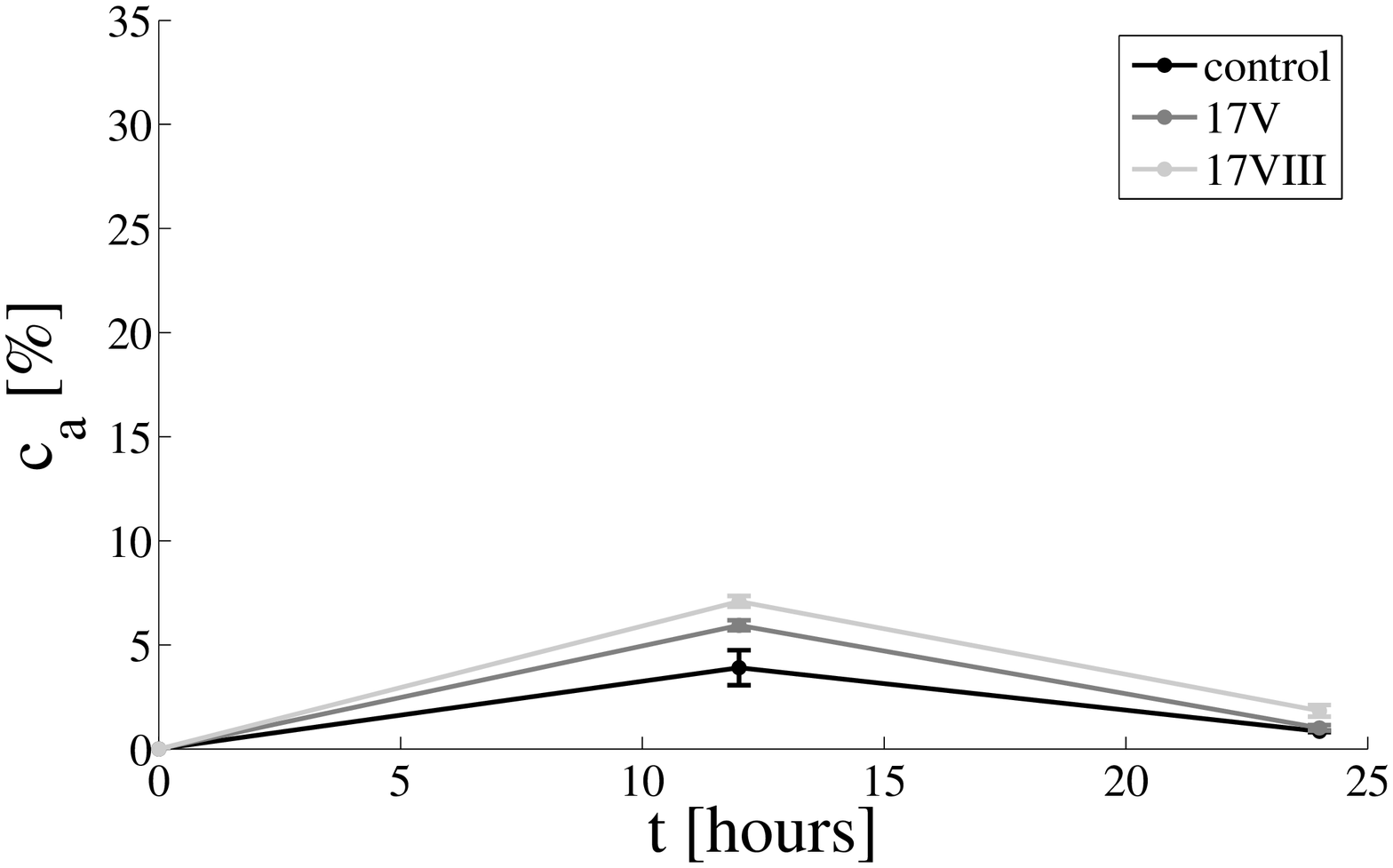,height=5.7cm}}}
\put(-4,6.0){$(a)$}
\end{picture}\par
\end{minipage}
\hfill
\begin{minipage}[t]{8.cm}
\begin{picture}(8.,8.)
\centerline 
{\hbox{\psfig{file=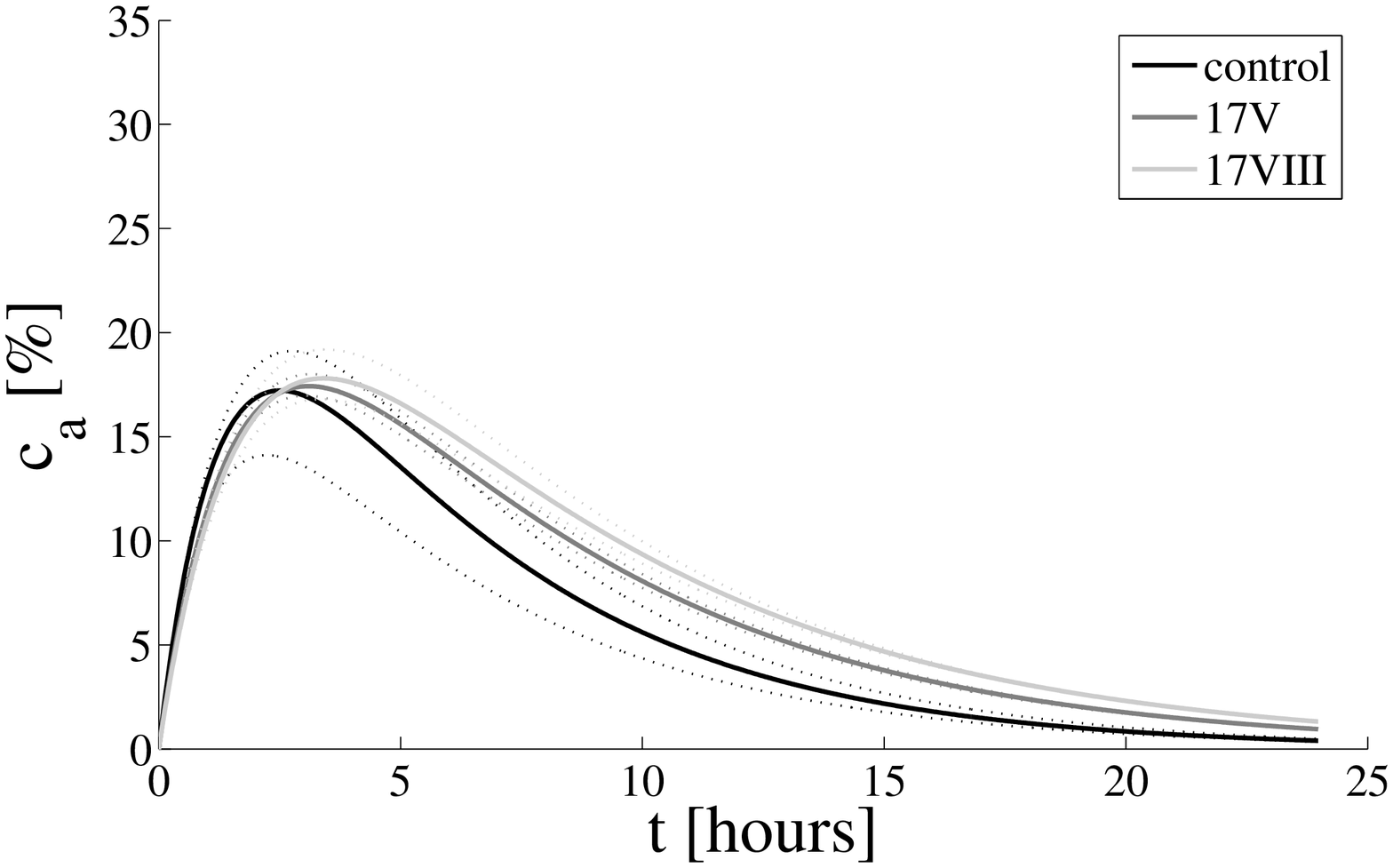,height=5.7cm}}}
\put(-4,6.0){$(b)$}
\end{picture}\par
\end{minipage}
\end{center}
\caption{\small \baselineskip=4pt {
{\bf Number of apoptotic cells after treatment with staurosporine $1\mu M,$ 
in the presence and in the absence of OS-9, 
from FACS analysis.}
}}
\label{fig:data:ca:5c:1muM}
\end{figure}

We then solve for $c_{a}(t)$ for [staurosporine]=$1\mu M,$ as given by
Eqn.~(\ref{eqn:ca}), and fit it to
Fig.~\ref{fig:data:ca:5c:1muM}(a) 
in order to determine the parameters $\gamma$ and $\delta.$ 
Setting $\beta, \delta =0$ and using $\alpha =0.20^{+0.005}_{-0.01},$ 
we fit the solution to the experimental curve for the number of
apoptotic cells in the negative control population and obtain 
$\gamma =0.72^{+0.18}_{-0.10}.$   
We proceed to fit the solution to the
experimental number of apoptotic cells in the OS-9-overexpressed cell line
populations, using the values determined above for the other
parameters, to find that 
$\delta_{17V} =0.19^{+0.17}_{-0.16}$  
and $\delta_{17VIII} =0.27^{+0.16}_{-0.13}.$  
The results are plotted in Fig.~\ref{fig:data:ca:5c:1muM}(b).


\subsubsection{[Staurosporine]=$0.1\mu M$}

\begin{figure}[t]
\setlength{\unitlength}{1cm}
\begin{center}
\begin{minipage}[t]{8.cm}
\begin{picture}(8.,8.)
\centerline 
{\hbox{\psfig{file=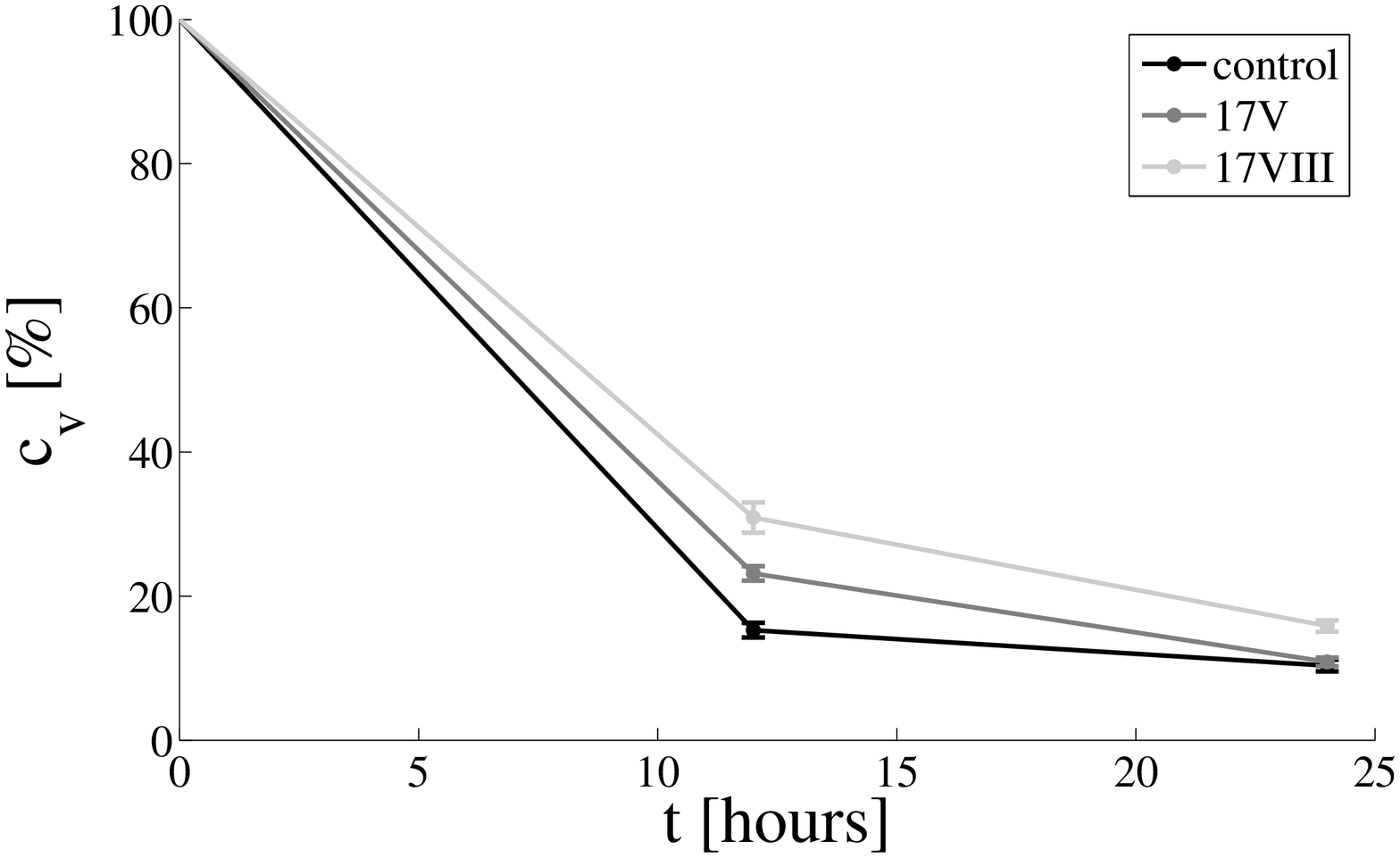,height=5.7cm}}}
\put(-4,6.0){$(a)$}
\end{picture}\par
\end{minipage}
\hfill
\begin{minipage}[t]{8.cm}
\begin{picture}(8.,8.)
\centerline 
{\hbox{\psfig{file=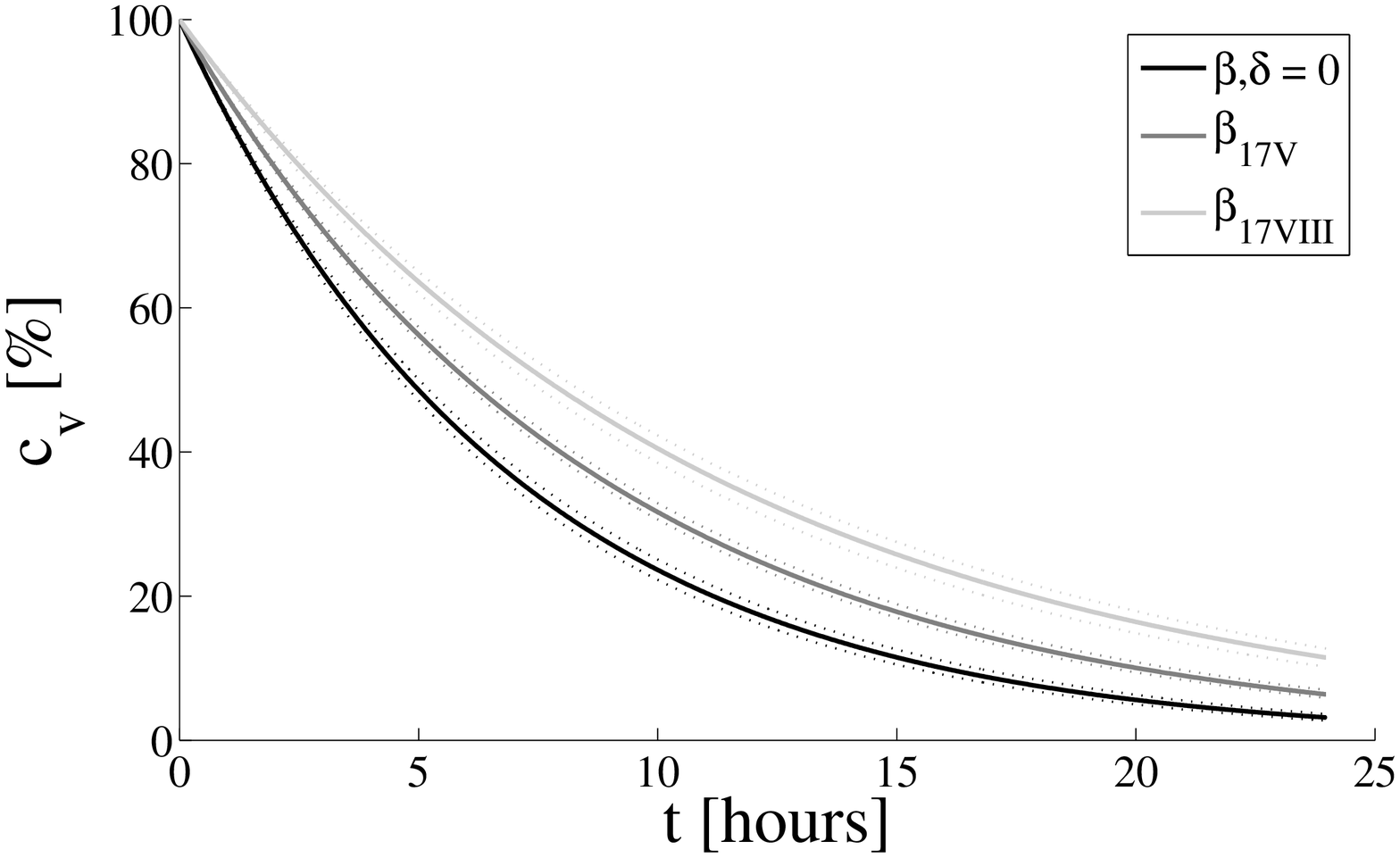,height=5.7cm}}}
\put(-4,6.0){$(b)$}
\end{picture}\par
\end{minipage}
\end{center}
\caption{\small \baselineskip=4pt {
{\bf Number of viable cells after treatment with
  staurosporine $0.1\mu M,$ in the presence and in the absence of OS9,
  from FACS analysis.}
}}
\label{fig:data:cv:5b:0.1muM}
\end{figure}

First we solve for $c_{v}(t),$ given by 
Eqn.~(\ref{eqn:cv}), and fit the solution
to the curves for $c_{v}$ depicted in Fig.~\ref{fig:data:cv:5b:0.1muM}(a).
Setting $\beta =0,$ we fit the solution to the experimental curve for
$c_v$ in the negative control population and find 
$\alpha =0.15^{+0.01}_{-0.005}.$ Fitting the solution to the
experimental curves for $c_v$ in the OS-9-overexpressed cell line
populations with $\alpha =0.15^{+0.01}_{-0.005}$ we find 
$\beta_{17V} =0.19^{+0.06}_{-0.05}$ and
$\beta_{17VIII} =0.35^{+0.05}_{-0.06}.$

\begin{figure}[p]
\setlength{\unitlength}{1cm}
\begin{center}
\begin{minipage}[t]{8.cm}
\begin{picture}(8.,8.)
\centerline 
{\hbox{\psfig{file=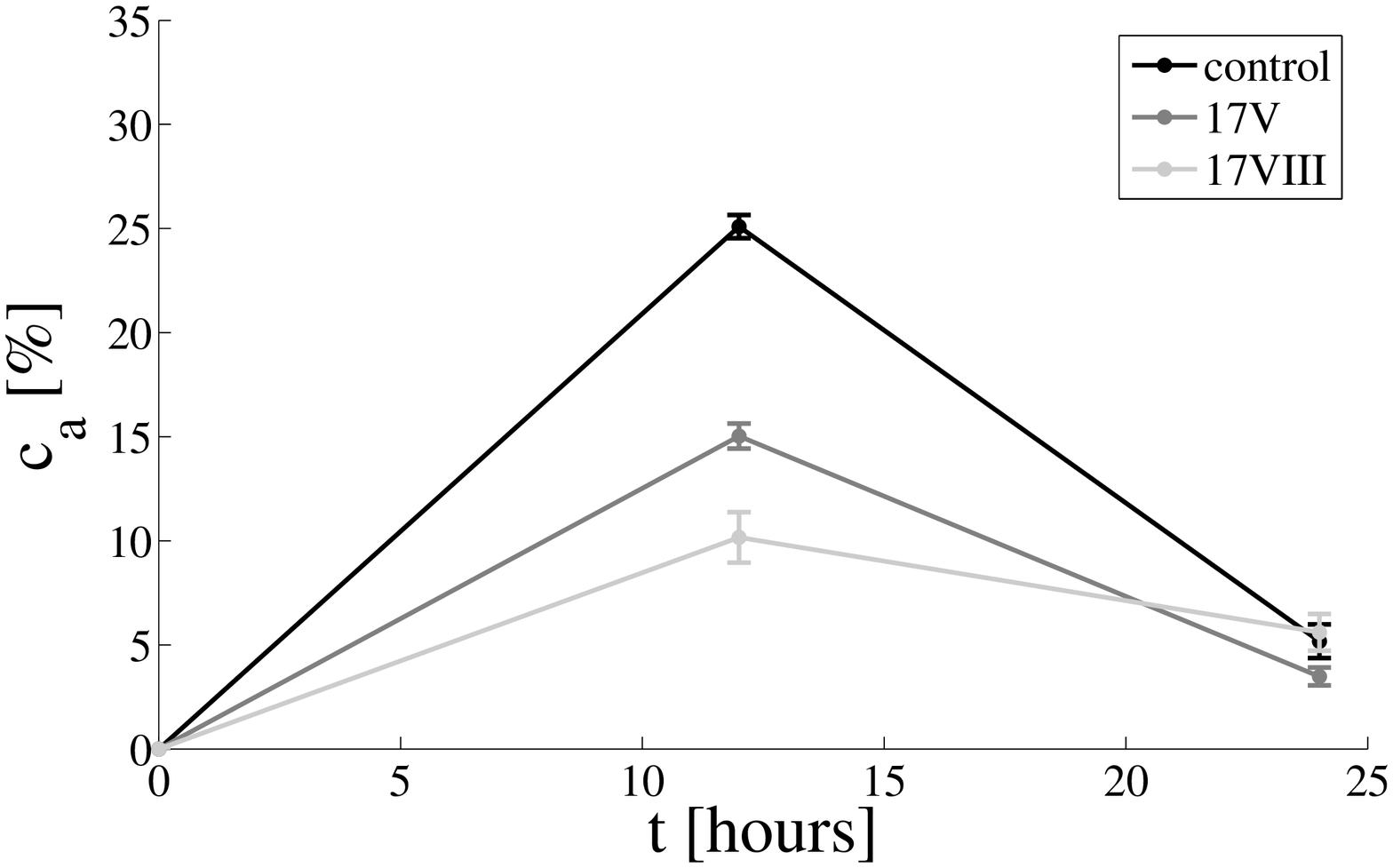,height=5.7cm}}}
\put(-4,6.0){$(a)$}
\end{picture}\par
\end{minipage}
\hfill
\begin{minipage}[t]{8.cm}
\begin{picture}(8.,8.)
\centerline 
{\hbox{\psfig{file=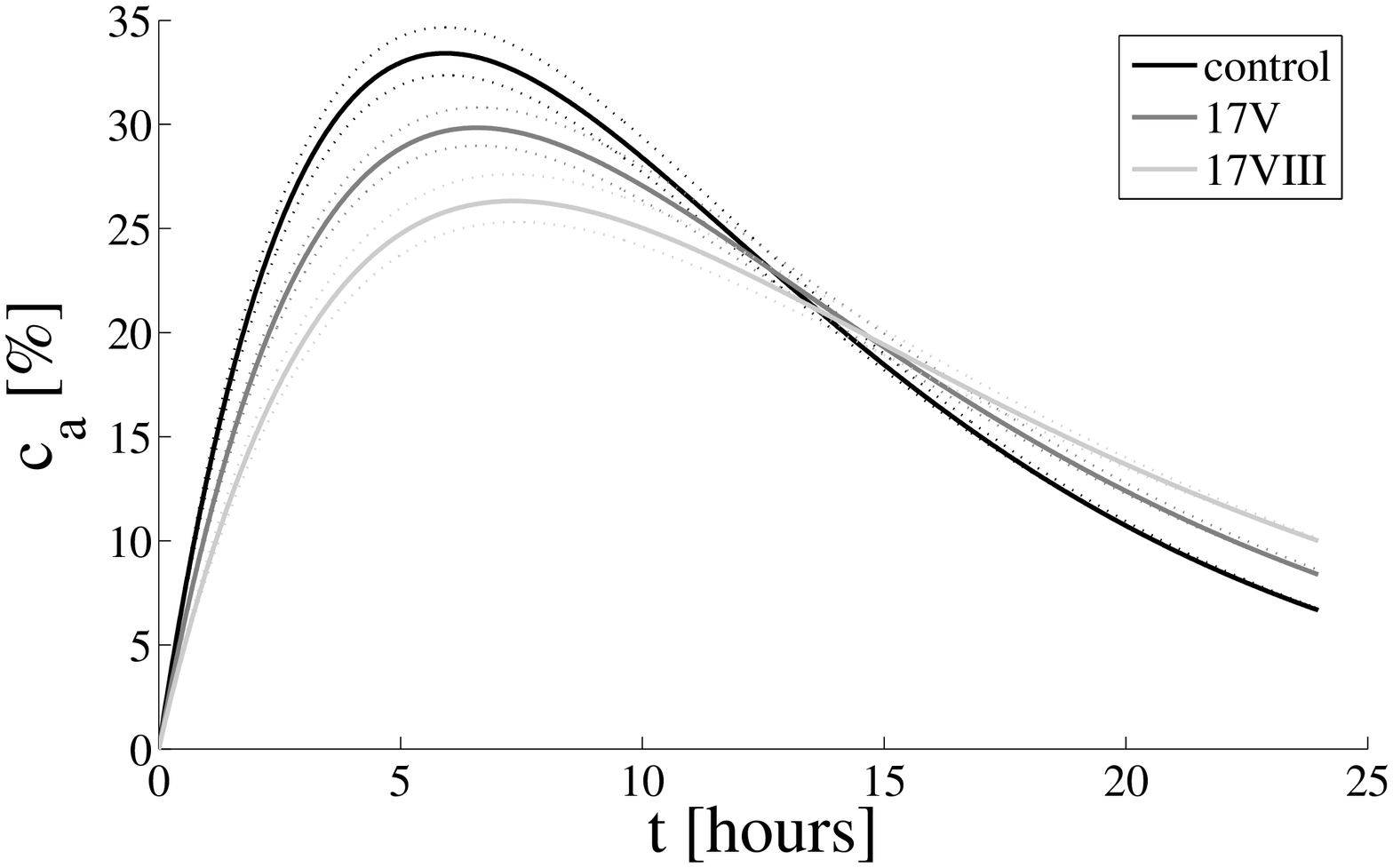,height=5.7cm}}}
\put(-4,6.0){$(b)$}
\end{picture}\par
\end{minipage}
\end{center}
\caption{\small \baselineskip=4pt {
{\bf Number of apoptotic cells after treatment with staurosporine $0.1\mu M,$ 
in the presence and in the absence of OS-9, 
from FACS analysis.}
}}
\label{fig:data:ca:5c:0.1muM}
\end{figure}

We then proceed to solve for $c_{a}(t),$ given by
Eqn.~(\ref{eqn:ca}), and fit the solution to the 
experimental curves for $c_{a}$ depicted in 
Fig.~\ref{fig:data:ca:5c:0.1muM}(a). 
Setting $\beta, \delta =0$ and $\alpha =0.15^{+0.01}_{-0.005},$ we fit the
solution to the experimental curve for $c_a$ in the negative control
population and find $\gamma =0.20^{+0.005}_{-0.01}.$ 
This value is substantially lower than the one obtained for
[staurosporine]=$1 \mu M$ ($\gamma =0.72^{+0.18}_{-0.10}$). This
difference emphasises the strong dependence of cell death rate with
the apoptotic agent.
Using now also
the values for $\beta,$ we fit the solution to the experimental curve
for $c_a$ in the 17V and 17VIII cell line populations to find both
$\delta_{17V} =0$ and $\delta_{17VIII} =0.$ 
These results are plotted in 
Fig.~\ref{fig:data:cv:5b:0.1muM}(b) and
Fig.~\ref{fig:data:ca:5c:0.1muM}(b). 


As anticipated by E.~Vourvouhaki et al. \cite{os9},
this case illustrates
resistance to apoptosis, primarily upstream ($\beta \not=0$) and
possibly also downstream
($\delta \not=0$), dependent on the concentration of staurosporine.  
This is supported by the  
observed dependence of the apoptosis time scale on the
concentration of the apoptotic agent. 
A higher concentration of staurosporine
increases the probability of interaction with a viable cell, 
as encapsulated in the larger value of $\alpha,$ 
$\alpha_{(1\mu M)} =0.20 >\alpha_{(0.1\mu M)}=0.15.$ 
The probability of survival 
is further reduced 
by the observed decrease of upstream resistance to apoptosis
with a higher concentration of staurosporine, 
which can be accounted for by the smaller value of $\beta,$ 
$\beta_{17V(1\mu M)}=0.18 <\beta_{17V(0.1\mu M)}=0.19$
and $\beta_{17VIII(1\mu M)}=0.24 <\beta_{17VIII(0.1\mu M)}=0.35.$
The observed higher dying rate of apoptotic cells, on the other hand,  
can be accounted for by the
larger decay rate $\gamma_{(1\mu M)}=0.72 >\gamma_{(0.1\mu M)}=0.20.$
The dependence of $\beta$ and $\gamma$ with the concentration of the
apoptotic agent was, however, not expected in the phenomenological
framework of the kinetical model. 


\subsection{TNF$\alpha$ treatment in FDC-P1 cells}

\begin{figure}[t]
\setlength{\unitlength}{1cm}
\begin{center}
\begin{minipage}[t]{8.cm}
\begin{picture}(8.,8.)
\centerline 
{\hbox{\psfig{file=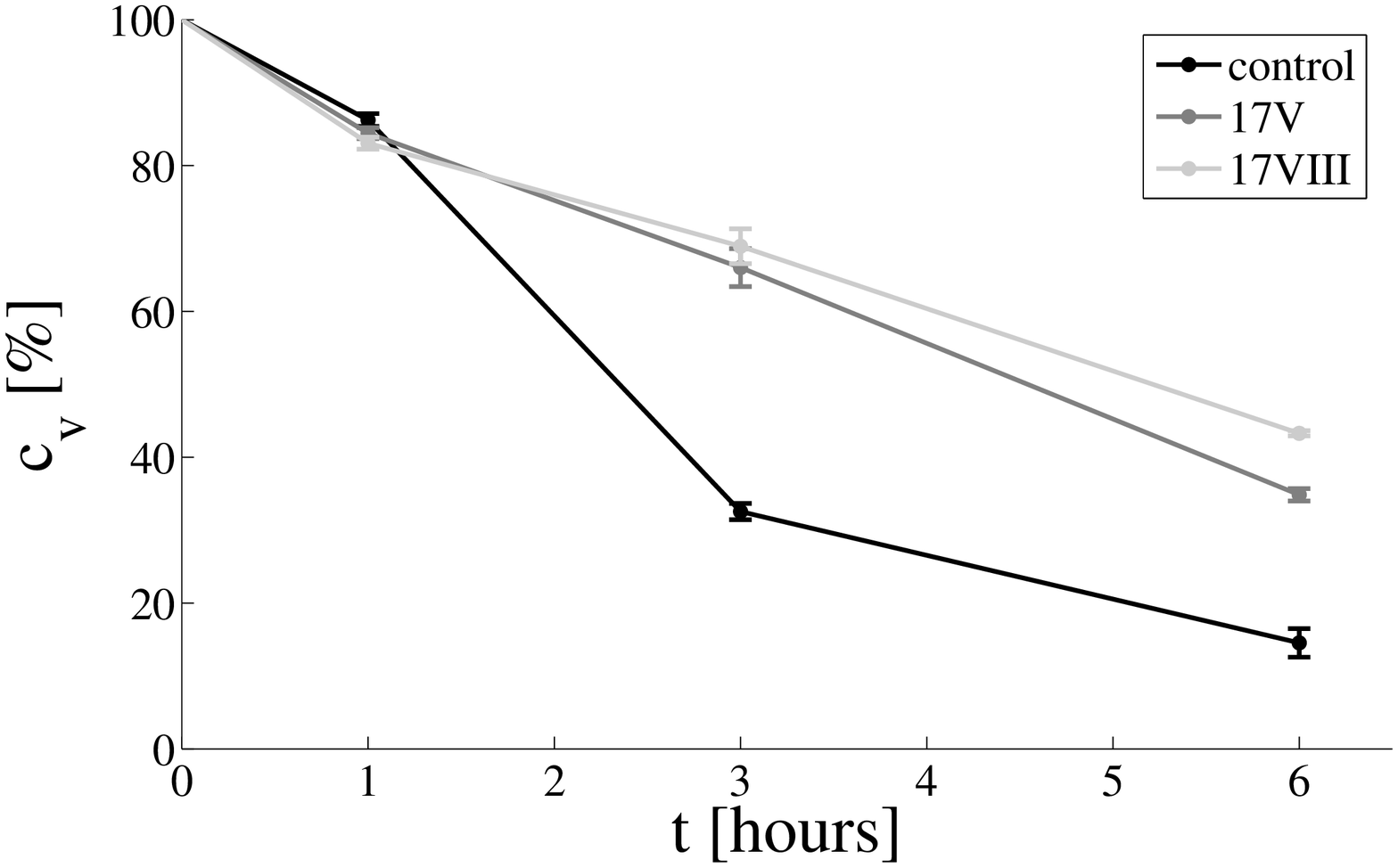,height=5.7cm}}}
\put(-4,6.0){$(a)$}
\end{picture}\par
\end{minipage}
\hfill
\begin{minipage}[t]{8.cm}
\begin{picture}(8.,8.)
\centerline 
{\hbox{\psfig{file=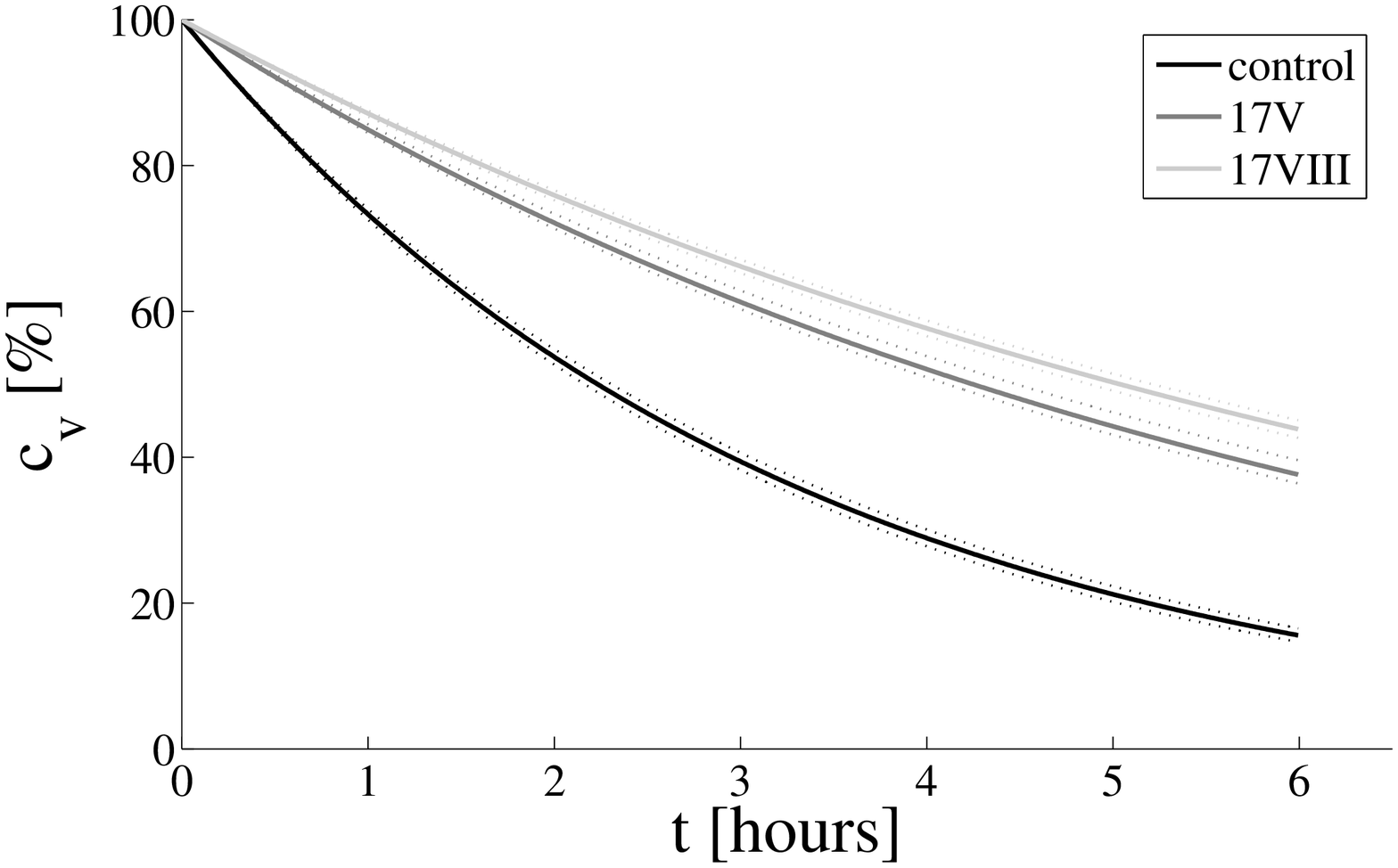,height=5.7cm}}}
\put(-4,6.0){$(b)$}
\end{picture}\par
\end{minipage}
\end{center}
\caption{\small \baselineskip=4pt {
{\bf Number of viable cells after treatment with TNF$\alpha,$ 
in the presence and in the absence of OS-9, 
from the FACS analysis.}
}}
\label{fig:data:cv:6b}
\end{figure}

\begin{figure}[p]
\setlength{\unitlength}{1cm}
\begin{center}
\begin{minipage}[t]{8.cm}
\begin{picture}(8.,8.)
\centerline 
{\hbox{\psfig{file=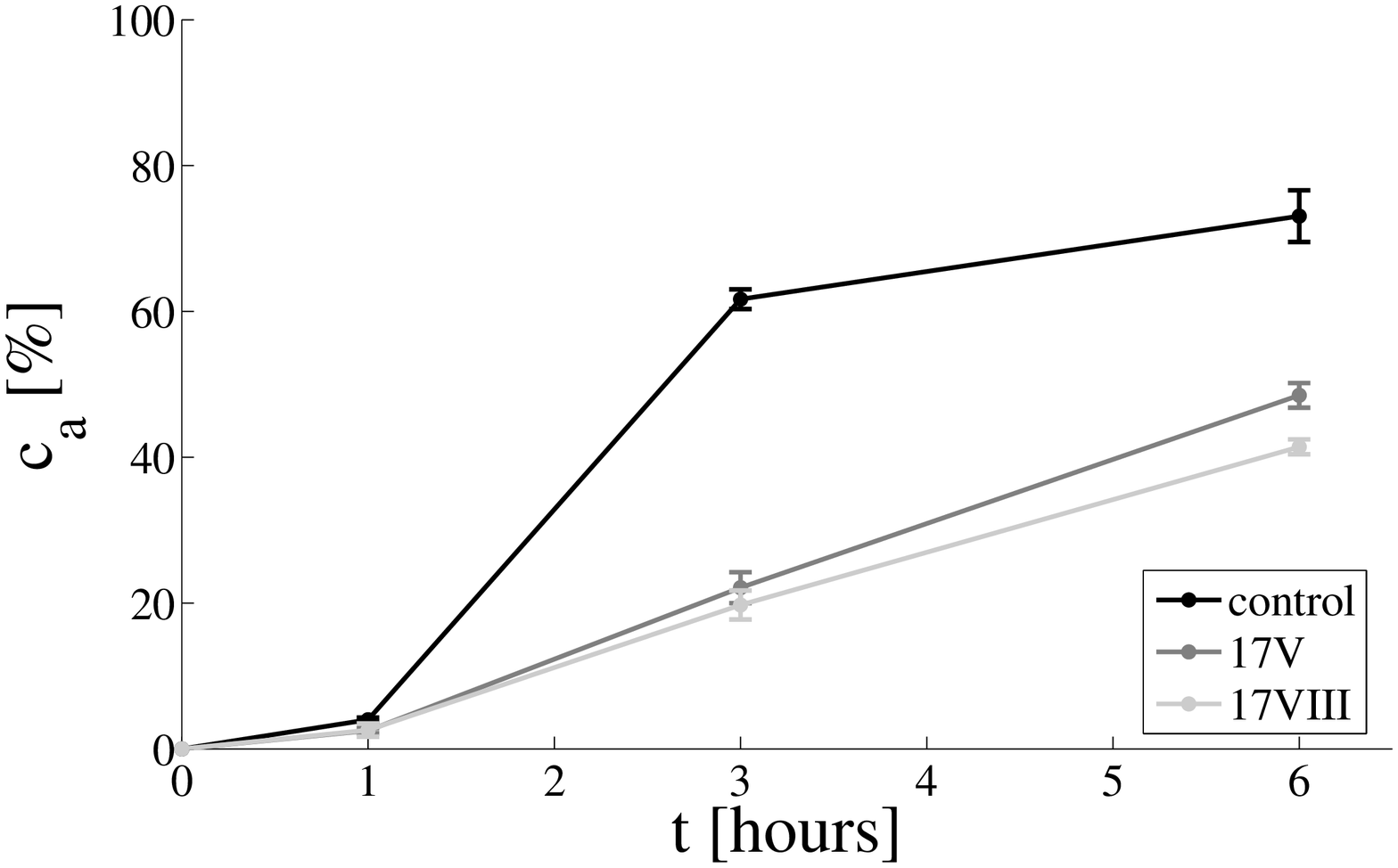,height=5.7cm}}}
\put(-4,6.0){$(a)$}
\end{picture}\par
\end{minipage}
\hfill
\begin{minipage}[t]{8.cm}
\begin{picture}(8.,8.)
\centerline 
{\hbox{\psfig{file=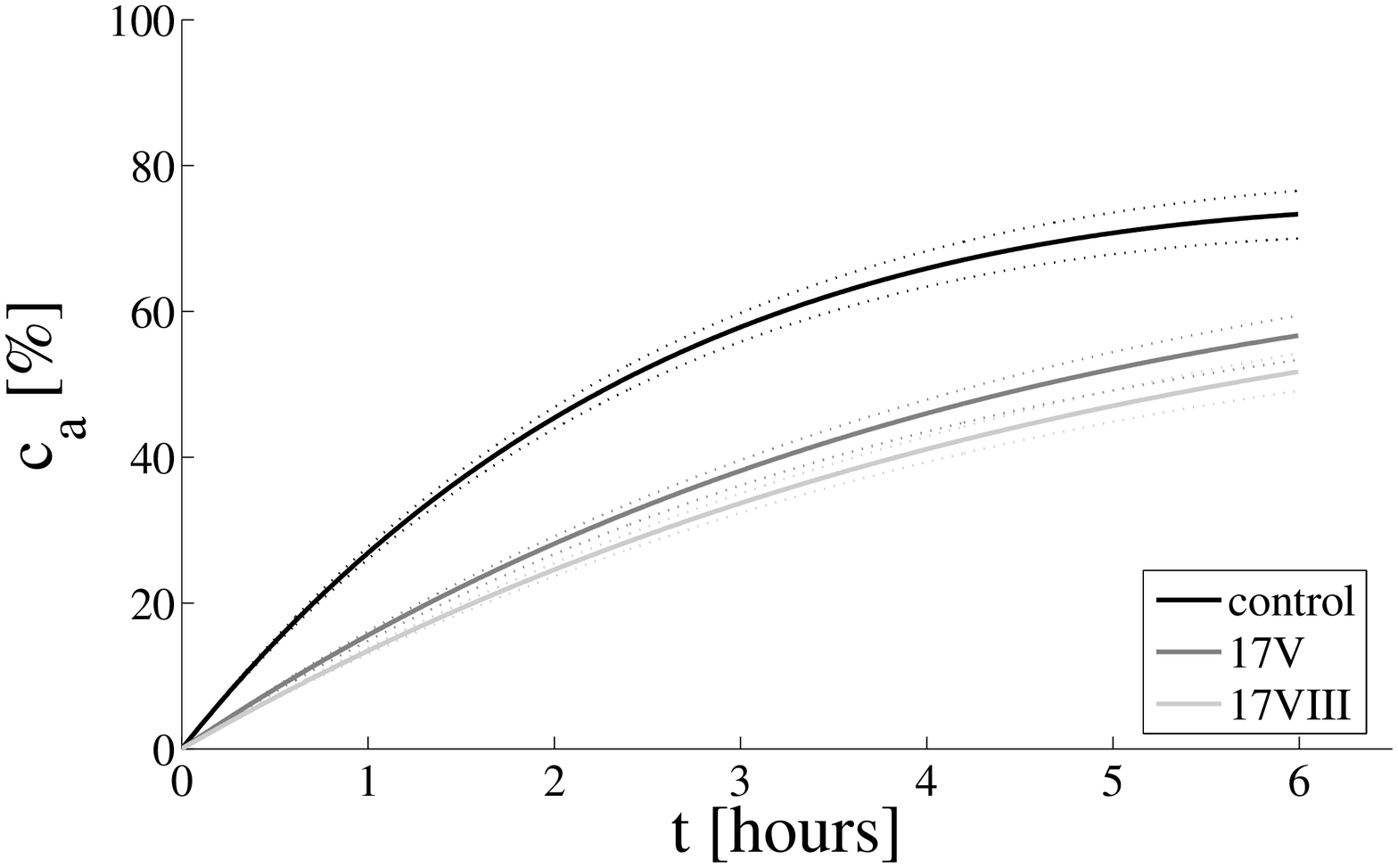,height=5.7cm}}}
\put(-4,6.0){$(b)$}
\end{picture}\par
\end{minipage}
\end{center}
\caption{\small \baselineskip=4pt {
{\bf Number of apoptotic cells after treatment with TNF$\alpha,$ 
in the presence and in the absence of OS-9,
from the FACS analysis.}
}}
\label{fig:data:ca:6c}
\end{figure}

Here we analyse  
apoptosis induction by treatment with TNF$\alpha$ (40ng/ml=$0.78\mu M$).
Similarly to the previous case discussed, 
the apoptotic mechanism is that of addition of an apoptotic agent,
which is described by the system of equations
(\ref{eqn:cv}--\ref{eqn:cn}), with $\beta$ having the interpretation
of OS-9 overexpression modulation in apoptosis resistance.  
The experimental results are collected 
in Figs.~6B and 6C of Ref.~\cite{os9}, which we
reproduce in Figs.~\ref{fig:data:cv:6b}(a) and
\ref{fig:data:ca:6c}(a) respectively. 

We want to solve for $c_{v}(t),$ given by
Eqn.~(\ref{eqn:cv}), 
and fit it to Fig.~\ref{fig:data:cv:6b}(a). 
Setting $\beta =0$, we fit the solution to the
experimental curve for $c_{v}$ in the negative control population cell
and find $\alpha =0.32^{+0.01}_{-0.01}.$  
Given $\alpha,$ we consider the case where
$\beta \not=0.$ Fitting the solution to the experimental curve
for $c_{v}$ in the OS-9-overexpressed cell line populations, we find 
$\beta_{17V} =0.46^{+0.04}_{-0.04}$  
and $\beta_{17VIII} =0.54^{+0.03}_{-0.03}.$ 
The resulting plots are shown in Fig.~\ref{fig:data:cv:6b}(b).

We then proceed to solve for $c_{a}(t),$ given by
Eqn.~(\ref{eqn:ca}), and to fit it to
Fig.~\ref{fig:data:ca:6c}(a). Using the values for $\alpha$ and
$\beta$ determined above, 
we fit the solution to the experimental curves for $c_{a}$ in the negative
control cell population and in the OS-9-overexpressed cell line
populations, finding 
respectively $\gamma =0.045^{+0.016}_{-0.016}$ 
and both $\delta_{17V} =0$ and $\delta_{17VIII} =0.$ 
The resulting plots are shown in
Fig.~\ref{fig:data:ca:6c}(b). 


From the parameter estimation we conclude that this case 
illustrates upstream resistance to apoptosis ($\beta \not=0$), in
agreement with what was anticipated by E.~Vourvouhaki et al.~\cite{os9}.

\section{Discussion}
\label{sec:discussion}

In this paper we present a simple theoretical model to assess the possible 
effects of OS-9 overexpression upon induction of apoptosis. 
We formulated a mechanistic model for the 
reaction of OS-9-overexpressed cell populations to induction of
apoptosis by different mechanisms and estimated the parameters of the
model from the data set available on the possible implication of OS-9
in promotion of cell viability and resistance to apoptosis.  
We summarize the results in Table~\ref{table:results}. \\

\begin{table}[h]
\begin{tabular}{|c|c|c|c|c|c|c|}
\hline
&$\alpha$ &$\beta_{17V}$ &$\beta_{17VIII}$
 &$\gamma$ &$\delta_{17V}$ &$\delta_{17VIII}$\\\hline
IL-3 
&$0.057^{+0.001}_{-0.001}$ &$0.17^{+0.04}_{-0.05}$ &$0.31^{+0.03}_{-0.04}$ 
&$0.30^{+0.04}_{-0.04}$ &$0.099^{+0.16}_{-0.099}$ &$0.086^{+0.14}_{-0.086}$\\  
\hline
staurosporine $1\mu M$ 
&$0.20^{+0.005}_{-0.01}$ &$0.18^{+0.005}_{-0.05}$ &$0.24^{+0.03}_{-0.07}$ 
&$0.72^{+0.18}_{-0.10}$ &$0.19^{+0.17}_{-0.16}$ &$0.27^{+0.16}_{-0.13}$\\
\hline
staurosporine $0.1\mu M$ 
&$0.15^{+0.01}_{-0.005}$ &$0.19^{+0.06}_{-0.05}$  &$0.35^{+0.05}_{-0.06}$ 
 &$0.20^{+0.005}_{-0.01}$ &$0$ &$0$\\ 
\hline
TNF$\alpha$ 
&$0.32^{+0.01}_{-0.01}$ &$0.46^{+0.04}_{-0.04}$ &$0.54^{+0.03}_{-0.03}$ 
&$0.045^{+0.016}_{-0.016}$ &$0$ &$0$\\
\hline
\end{tabular}
\caption{\small \baselineskip=4pt {
{\bf Kinetical model parameters estimated for both cell
line populations.}
}}
\label{table:results}
\end{table}

From this Table we observe that the OS-9 overexpression-dependent
parameters take non-vanishing values. Moreover, comparing the $\beta$
values across all the apoptotic inducing mechanisms, we observe a
strong dependence with the level of OS-9 overexpression in the two
cell line populations 17V and 17VIII. 
These observations suggest that OS-9 has a possible role 
as an anti-apoptotic factor  
and that this role is dose-dependent, with the cell line population
17VIII exhibiting a greater effect in both promoting cell viability
and resisting to induction of apoptosis than 17V.

A closer look at the $\delta$ values reveals that they are very
close to zero except for one situation, namely [staurosporine]
$=1\mu M$ for the 17VIII cell line population. We are persuaded to
speculate that the OS-9 overexpression does not have a significant
effect on downstream modulation of apoptosis.
The vanishing values found for the parameter $\delta$ whenever
  presented without error bars are strictly vanishing, i.e. in the
  parameter interval $[0, 1]$ the minimum of the cost function is at
  $\delta=0.$ In the cases where the {\it induced average value} for
  $\delta$ is non-vanishing, the {\it induced error bars} can still
  contemplate a vanishing $\delta.$ 
A possible way of testing this effect would be  
to measure the OS-9 overexpression in the apoptotic phase only, as
described by Eqn.~(\ref{death}). The number of apoptotic cells
would be controlled, so that the conversion rate will only depend on
$\gamma$ and $\delta.$






We observe that our model reflects the tendency observed in the data
collected in Ref.~\cite{os9} and supports the interpretation therein
advanced. There is quantitative evidence that OS-9 has a role in the
regulation of the conversion rates for $c_{v}\to c_{a}$ and $c_{a}\to
c_{n},$ as illustrated by the non-vanishing values found for the
OS-9-dependent parameters.
Such evidence, however, is not determinant since the values are small
and the errors are large (ranging from around 
10-100\% of the values estimated from the average data values).
The results show that OS-9 contributes primarily to promotion of
viability and upstream resistance to apoptosis, as reflected by
$\beta \not=0,$ with little or no effect on downstream resistance, as
reflected by the close to vanishing values for $\delta.$
Furthermore, this function does not appear to be related to an increase
in cell proliferation, as DNA content analysis shows that cells
overexpressing OS9 do not have a greater proportion of dividing
cells than control populations \cite{os9}.


Some results, however, remain to be fully explained.
We collapsed in the parameter $\alpha$ all the intricacies of the
  intra-cellular, molecular processes involved in apoptosis
  induction and whose relation with OS-9 we had no data to constrain
  or infer.
TNF$\alpha$ gives us the highest rate of apoptosis induction among the
various other factors, followed by [staurosporine]$=1\mu M$ and
[staurosporine]$=0.1\mu M.$
The withdrawal of IL-3, being a factor that promotes cell viability,
can be seen indirectly to affect cell viability. 
Of the three apoptosis induction mechanisms described in the
  available data sets, only the staurosporine-mediated one allowed to
  discriminate the 
  effect of the concentration of the apoptotic agent, $[a_a].$ For that data
  set we observe a dependence of $\alpha$ with $[a_a].$ However, our
  model cannot predict the functional form of this dependence so as to
  reproduce the value $A=0.125^{+0.003}_{-0.005}$ in the empirical expression
\ba
{\alpha _{(1\mu M)}\over \alpha _{(0.1\mu M)}} 
=\left( {1\over 0.1}\right)^{A}
\ea
relating the ten-fold increase of $[a_a]$ with $\alpha.$
Moreover, we observe a dependence of both $\beta$ and $\gamma$ with
$[a_a]$ which
is not {\it a priori} encompassed by the kinetical
model, since these parameters measure the rate of cellular processes whose
developement has, to our current understanding, no relation to 
the environmental conditions that triggered the preceding process.
We also note that both $\beta$ and $\gamma$ act towards
a decrease of $c_a,$ though in different ways. We venture that these
observations might be evidence of the limits of the descriptive power of the
kinetical model in face of the complexity of the biological system
that it was conceived to describe as a first-order model.
The binding rates $\alpha _{up (i)}$ and $\alpha _{down (i)}$ in the chemical
model conceal intermediate phenomena at levels of interaction
unaccounted for by the kinetical model. 
We believe that a higher-order model would be able to expose the
  sources of these observations, as well as allow for possible
  dependencies with the apoptotic agent or the apoptosis inducing mechanism.

Additional data from various experimental setups and from
complementary measurements are, therefore, required to
unambiguously determine the function of OS-9 in cell fate and
in tumour growth, 
and test the model as a realistic theoretical description of the
biological system. 
In particular, further experiments should attempt to identify the
signalling pathway through which OS-9 exerts its function.
Biochemical assays and western blot analysis will provide the means
for the identification of the proteins that OS-9 binds to in order
to transmit its potential signals. It is also essential that the function
of OS-9 should be tested in {\it in vivo} models. The establishment of
knockout mice is also important in order to assess the functional
importance of OS-9 during development.
Only then can we rely on the values for the parameters
estimated by the mathematical technique developed in this paper. 


Addition of an apoptotic agent or withdrawal of a vital agent are expected
to induce apoptosis
through different pathways, respectively by a caspase-initiated cascade 
\cite{staurosporine1, staurosporine2, tnfalpha} 
or by the mitochondria/cytochrome c pathway
 \cite{il3}. 
We note, however, that we cannot compare the concentrations
of different reagents/proteins 
since each one of them
acts in a different way in the cells. 
Different molecular processes entailing different cellular mechanisms
can nevertheless reflect the same populational effect, thus rendering
themselves to be described by the same mathematical model. 
Though at the level of description of our first-order model the
parameters estimated for each mechanism are not comparable,
they are nonetheless valid to suggest a
contribution of OS-9 to cell viability and resistance to apoptosis.

The process of explaining a mechanism involves an active process of
   simplification: it is necessary to make assumptions to define the
   most relevant components contributing to the overall 
   observed dynamics, while removing factors which contribute less. 
The isolation and identification of the core
   principles governing the dynamics is paramount to the detailed
   understanding of any system. [For a discussion, see
   Ref.~\cite{ref}].
Following this approach, we present
   a minimal model which is capable of capturing the structure of
   the experimental data. This model does not fit perfectly the
   experimental data but it is built upon biologically plausible and
   consistent assumptions. The values for all the parameters in the
   model are calculated using the least mean square method and are,
   in that sense, optimal. More complex models could be derived which,
   by using more parameters, would provide a better fit to the
   data. However, such models would be less informative since the
   functional role of the additional parameters would be
   speculative. Moreover, experimental data are always noisy and a
   perfect fit to a specific data set means loss in generality. It is
   important to emphasise that very little is known about the dynamics
   associated with OS-9. Our minimal model conciliates the scarce
   information available with robust assumptions to produce a simple
   but general model capable of quantifying the contribution of OS-9 as
   a potent anti-apoptotic factor. This is a first-order model which
   serves as a valuable starting point for experiments aiming at a
   more detailed understanding of the role of OS-9. To the best of
   our knowledge, this is the only available model for the
   quantification of the OS-9 contribution to apoptosis.

This model can be applied to other cellular systems since it allows for the
discrimination of effects at different stages in a complex chemical
mechanism and can easily be upgraded in accordance to the functional
information available. 

\vskip 0.5cm

{\bf Acknowledgments}

The authors dedicate this paper to the memory of Peny Tziamourani.
C.C. and P.A. thank the Funda\c{c}\~ao para a Ci\^encia e a Tecnologia
for support. C.C. also thanks the National and Kapodistrian
University of Athens for its hospitality.

\newpage

\section{Captions}

Figures denoted by (a) reproduce the data reported in Ref.~\cite{os9},
the points depicted giving the readings for the different cell
populations as follows:  
HA-expressed cell population (black),
17V cell line (dark gray),
17VIII cell line (light gray).
Figures denoted by (b) show the solutions for either $c_v(t)$ or
$c_a(t).$ \\

\noindent{\bf Fig.~\ref{fig:data:cv:2b}: 
Number of viable cells after withdrawal of IL-3,
in the presence and in the absence of OS-9, 
from FACS analysis.}
In (a) we reproduce the data from Fig.~2B in Ref.~\cite{os9}.
In (b) we show the solutions
for $c_{v}(t)$ with $\alpha =0.057.$
The black curve depicts $c_{v}(t)$ when $\beta=0;$ 
the gray curves depict $c_{v}(t),$ with the dark
gray curve for $\beta_{17V} =0.17$ and the light grey curve for
$\beta_{17VIII} =0.31.$  
(The upper dashed curves represent the fittings to the maxima data with
$\alpha =0.056,$ 
$\beta _{17V} =0.19$ and $\beta _{17VIII} =0.32;$ 
the lower dashed curves represent the fittings to the minima data with
$\alpha_{0} =0.058,$ 
$\beta _{17V} =0.15$ and $\beta _{17VIII} =0.30.$)
\vskip 0.5cm

\noindent{\bf Fig.~\ref{fig:data:ca:2c}
Number of apoptotic cells after withdrawal of IL-3, 
in the presence and in the absence of OS-9, 
from FACS analysis.}
In (a) we reproduce the data from Fig.~2C in Ref.~\cite{os9}.
In (b) we show the solutions
for $c_{a}(t)$ with $\alpha =0.057$ and $\gamma =0.30.$ 
The black curve depicts $c_{a}(t)$ when $\beta, \delta =0;$ 
the gray curves depict $c_{a}(t),$ with the dark
gray curve for $\beta_{17V} =0.17$ and $\delta_{17V} =0.099,$
and the light gray curve for
$\beta_{17VIII} =0.31$ and $\delta_{17VIII} =0.086.$
(The upper dashed curves represent the fittings to the maxima data with
$\alpha =0.058,$ $\gamma =0.26,$ 
$\beta _{17V} =0.15$ and $\delta_{17V} =0.070,$ 
and $\beta _{17VIII} =0.30$ and $\delta_{17VIII} =0.029;$ 
the lower dashed curves represent the fittings to the minima data with
$\alpha =0.056,$ $\gamma =0.34,$ 
$\beta _{17V} =0.19$ and $\delta_{17V} =0.14,$ 
and $\beta _{17VIII} =0.32$ and $\delta_{17VIII} =0.16.$) 

\vskip 0.5cm

\noindent{\bf Fig.~\ref{fig:data:cv:5b:1muM}
Number of viable cells after treatment with
  staurosporine $1\mu M,$ in the presence and in the absence of OS9,
  from FACS analysis.}
In (a) we reproduce the data from Fig.~5B in Ref.~\cite{os9}.
In (b) we show the solutions
for $c_{v}(t)$ with $\alpha =0.20.$
The black curve depicts $c_{v}(t)$ when $\beta =0.$ 
The gray curves depict $c_{v}(t),$ with the dark gray
curve for $\beta_{17V} =0.18$ and the light gray curve for
$\beta_{17VIII} =0.24.$  
(The upper dashed curves represent the fittings to the maxima data with
$\alpha =0.19,$ 
$\beta _{17V} =0.14$ and $\beta _{17VIII} =0.23;$ 
the lower dashed curves represent the fittings to the minima data with
$\alpha =0.20,$ 
$\beta _{17V} =0.17$ and $\beta _{17VIII} =0.20.$)

\vskip 0.5cm

\noindent{\bf Fig.~\ref{fig:data:ca:5c:1muM}
Number of apoptotic cells after treatment with staurosporine $1\mu M,$ 
in the presence and in the absence of OS-9, 
from FACS analysis.}
In (a) we reproduce the data from Fig.~5C in Ref.~\cite{os9}.
In (b) we show the solutions
for $c_{a}(t)$ with $\alpha =0.20$ and $\gamma =0.72.$ 
The black curve depicts $c_{a}(t)$ when $\beta, \delta =0;$ 
the gray curves depict $c_{a}(t),$ with the dark
gray curve for $\beta_{17V} =0.18$ and $\delta_{17V} =0.19,$
and the light gray curve for
$\beta_{17VIII} =0.24$ and $\delta_{17VIII} =0.27.$
(The upper dashed curves represent the fittings to the maxima data with
$\alpha =0.20,$ $\gamma =0.62,$ 
$\beta _{17V} =0.17$ and $\delta_{17V} =0.09,$ 
and $\beta _{17VIII} =0.20$ and $\delta_{17VIII} =0.20;$ 
the lower dashed curves represent the fittings to the minima data with
$\alpha =0.19,$ $\gamma =0.90,$ 
$\beta _{17V} =0.14$ and $\delta_{17V} =0.33,$ 
and $\beta _{17VIII} =0.23$ and $\delta_{17VIII} =0.39.$) 

\vskip 0.5cm

\noindent{\bf Fig.~\ref{fig:data:cv:5b:0.1muM}
Number of viable cells after treatment with
  staurosporine $0.1\mu M,$ in the presence and in the absence of OS9,
  from FACS analysis.}
In (a) we reproduce the data from Fig.~5B in Ref.~\cite{os9}.
In (b) we show the solutions 
for $c_{v}(t)$ with $\alpha =0.15.$
The black curve depicts $c_{v}(t)$ when $\beta =0;$
the gray curves depict $c_{v}(t),$ with
the dark gray curve giving $\beta_{17V} =0.19$ and
the light gray curve $\beta_{17VIII} =0.35.$
(The upper dashed curves represent the fittings to the maxima data with
$\alpha =0.15,$ 
$\beta _{17V} =0.18$ and $\beta _{17VIII} =0.35;$ 
the lower dashed curves represent the fittings to the minima data with
$\alpha =0.16,$ $\beta _{17V} =0.20$ and $\beta _{17VIII} =0.34.$)

\vskip 0.5cm

\noindent{\bf Fig.~\ref{fig:data:ca:5c:0.1muM}
Number of apoptotic cells after treatment with staurosporine $0.1\mu M,$ 
in the presence and in the absence of OS-9, 
from FACS analysis.}
In (a) we reproduce the data from Fig.~5C in Ref.~\cite{os9}.
In (b) we show the solutions
for $c_{a}(t)$ with $\alpha =0.15$ and  $\gamma =0.20.$  
The black curve depicts $c_{a}(t)$ when $\beta, \delta =0;$ 
the gray curves depict $c_{a}(t),$ 
with $\beta_{17V} =0.19$ and $\delta_{17V} =0$ for the dark gray curve, 
with $\beta_{17VIII} =0.35$ and $\delta_{17VIII} =0$ for the light gray curve.
(The upper dashed curves represent the fittings to the maxima data with
$\alpha =0.16,$ $\gamma =0.19,$ 
$\beta _{17V} =0.20$ and $\delta_{17V} =0,$ 
and $\beta _{17VIII} =0.34$ and $\delta_{17VIII} =0;$ 
the lower dashed curves represent the fittings to the minima data with
$\alpha =0.15,$ $\gamma =0.20,$ 
$\beta _{17V} =0.18$ and $\delta_{17V} =0,$ 
and $\beta _{17VIII} =0.35$ and $\delta_{17VIII} =0.$) 

\vskip 0.5cm

\noindent{\bf Fig.~\ref{fig:data:cv:6b}
Number of viable cells after treatment with TNF$\alpha,$ 
in the presence and in the absence of OS-9, 
from the FACS analysis.}
In (a) we reproduce the data from Fig.~6B in Ref.~\cite{os9}.
In (b) we show the solutions
for $c_{v}(t)$ with $\alpha =0.32.$
The black curve depicts $c_{v}(t)$ when $\beta=0;$ 
the gray curves depict $c_{v}(t),$ with the dark
gray curve for $\beta_{17V} =0.46$ and the light grey curve for
$\beta_{17VIII} =0.54.$  
(The upper dashed curves represent the fittings to the maxima data with
$\alpha_{0} =0.31,$ 
$\beta _{17V} =0.47$ and $\beta _{17VIII} =0.54;$ 
the lower dashed curves represent the fittings to the minima data with
$\alpha_{0} =0.33,$ 
$\beta _{17V} =0.46$ and $\beta _{17VIII} =0.54.$)

\vskip 0.5cm

\noindent{\bf Fig.~\ref{fig:data:ca:6c}
Number of apoptotic cells after treatment with TNF$\alpha,$ 
in the presence and in the absence of OS-9,
from the FACS analysis.}
In (a) we reproduce the data from Fig.~6C in Ref.~\cite{os9}.
In (b) we show the solutions
for $c_{a}(t)$ with $\alpha =0.32$ and $\gamma =0.045.$ 
The black curve depicts $c_{a}(t)$ when $\beta, \delta =0;$ 
the gray curves depict $c_{a}(t),$ with the dark
gray curve for $\beta_{17V} =0.46$ and $\delta_{17V} =0,$
and the light gray curve for
$\beta_{17VIII} =0.54$ and $\delta_{17VIII} =0.$
(The upper dashed curves represent the fittings to the maxima data with
$\alpha =0.33,$ $\gamma =0.036,$ 
$\beta _{17V} =0.46$ and $\delta_{17V} =0,$ 
and $\beta _{17VIII} =0.54$ and $\delta_{17VIII} =0;$ 
the lower dashed curves represent the fittings to the minima data with
$\alpha =0.31,$ $\gamma =0.055,$ 
$\beta _{17V} =0.47$ and $\delta_{17V} =0.33,$ 
and $\beta _{17VIII} =0.54$ and $\delta_{17VIII} =0.$) \\

\noindent{\bf Table ~\ref{table:results}: 
Kinetical model parameters estimated for both cell
 line populations.} We summarize the values of the parameters estimated
for both cell line populations and for the three different mechanisms of
apoptosis induction.

\clearpage

\section{Figures}

\newpage

\newpage

\newpage

\begin{thebibliography}{99}

\bibitem{1}
N.~Johnson, S.~Sengupta, S.A.~Saidi, K.~Lessen,S.D.~Charnock-Jones,
L.~Scott, R.~Stephens,T.C.~Freeman,B.D.~Tom, M.~Harris,G.~Denyer,
M.~Sundaram, R.~Sasisekharan~R., S.K.~Smith, C.G.~Print, 
Endothelial cells preparing to die by apoptosis initiate a program
of transcriptome and glycome regulation, FASEB J. {\bf 18}(1), 188-190 (2004).

\bibitem{8}
Y.~Kimura, M.~Nakazawa and M.~Yamada,
Cloning and characterization of three isoforms of OS-9 cDNA and
expression of the OS-9 gene in various human tumor cell lines,
J Biochem {\bf 123}(5), 876-882 (1998). 

\bibitem{11}
Y.~Kimura, M.~Nakazawa, N.~Tsuchiya, S.~Asakawa, N.~Shimizu and
M.~Yamada, 
Genomic organization of the OS-9 gene amplified in human sarcomas,
 J Biochem {\bf 122}(6), 1190-1195 (1997).

\bibitem{archenemy1}
J.H.~Baek, P.C.~Mahon, J.~Oh, B.~Kelly, B.~Krishnamachary, M.~Pearson,
D.A.~Chan, A.J.~Giaccia and G.L.~Semenza,
OS-9 interacts with hypoxia-inducible factor 1alpha and prolyl
hydroxylases to promote oxygen-dependent degradation of HIF-1alpha,
 Mol Cell. {\bf 17}(4), 503-512 (2005).

\bibitem{archenemy2}
N.~Vigneron, A.~Ooms, S.~Morel, G.~Degiovanni, B.J.~Van Den Eynde,
Identification of a new peptide recognized by autologous cytolytic T
lymphocytes on a human melanoma,
Cancer Immun. {\bf 2}, 9 (2002). 

\bibitem{os9}
E.~Vourvouhaki, S.M.~Sullivan, Y.~deNantois, S.K.~Smith, C.G.~Print and
D.S.~Charnock-Jones,
OS-9 as a potent anti-apoptotic factor that also promotes cell viability,
Rev Clin Pharmacol Parmacokinet (2007), (ahead of print). 

\bibitem{il3}
D.E.~Johnson, 
Regulation of survival pathways by IL-3 and induction of apoptosis
following IL-3 withdrawal, 
Front Biosci {\bf 3}, d313-324 (1998).

\bibitem{staurosporine1}
T.L.~Yue, C.~Wang, A.M.~Romanic, K.~Kikly, P.~ Keller, W.E.~DeWolf,
T.K.~Hart, H.C.~Thomas, B.~Storer, J.L.~Gu, X.~Wang and G.Z.~Feuerstein,
Staurosporine-induced apoptosis in cardiomyocytes: A potential role of
caspase-3, J Mol Cell Cardiol. {\bf 30}(3), 495-50 (1998).

\bibitem{staurosporine2}
G.~Feng and N.~Kaplowitz, Mechanism of staurosporine-induced apoptosis in
murine hepatocytes, Am J Physiol Gastrointest Liver Physiol {\bf 282},
G825-G834 (2002).

\bibitem{tnfalpha}
E.E.~Varfolomeev and A.~Ashkenazi, Tunor necrosis factor: an apoptosis
JuNKie?, Cell {\bf 116}, 491-497 (2004).


\bibitem{cbc} D.~Wodarz and N.L.~Komarova, 
Computational Biology of Cancer: lecture notes and mathematical
  modeling, World Scientific, 2005.

\bibitem{mm}
A.S.~Novozhilov, F.S.~Berezovskaya, E.V.~Koonin and G.P.~Karev,
Mathematical modelling of anti-tumor virus therapy: regimes with
complete recovery within the framework of deterministic models,
q-bio.TO/0512022. 

\bibitem{apoptosis}
B.~Zhivotovsky and S.~Orrenius, 
Defects in the apoptotic machinery of cancer cells: role in drug
resistance,
Seminars in Cancer Biology {\bf 13}, 125-134 (2003).



\bibitem{ref}
J.M.G.~Vilar, C.C.~Guet and S.~Leibler, Modeling network dynamics: the
{\it lac} operon, a case study, Cell Biol. {\bf 161}, 471 (2003).




\end{thebibliography}
\end{document}